\documentclass[a4paper,10pt]{article}
\usepackage{latexsym}
\usepackage{amsmath}
\usepackage{amssymb}
\usepackage{amscd}

\newlength{\minitwocolumn}
\makeatletter
\@addtoreset{equation}{section}
\makeatother

\title{Form factors 
of spin $1$ analogue \\ 
of the eight-vertex model}

\author{Yas-Hiro Quano\thanks
{email: quanoy@suzuka-u.ac.jp}}

\date{\it Department of Clinical Engineering, 
Suzuka University of Medical Science \\
      \it Kishioka-cho, Suzuka 510-0293, Japan \\
22 Sep 2015
}

\begin{document}

\maketitle
\begin{abstract}
The twenty-one-vertex model, 
the spin $1$ analogue of the eight-vertex model is considered 
on the basis of free field representations of vertex operators 
in the $2\times 2$-fold fusion SOS model and vertex-face transformation. 
The tail operators, which translate corner transfer matrices 
of the twenty-one-vertex model into those of the fusion SOS model, 
are constructed by using free bosons and fermions 
for both diagonal and off-diagonal matrix elements 
with respect to the ground state sectors. 
Form factors of any local operators are therefore obtained in terms of 
multiple integral formulae, in principle. 
As the simplest example, the two-particle form factor of the spin 
operator is calculated explicitly. 
\end{abstract}

\section{Introduction}

In this paper we consider the spin $1$ analogue of 
Baxter's eight-vertex model \cite{ESM}, 
on the basis of vertex operator approach \cite{JMbk}. 
The model is often called twenty-one-vertex model 
since the $R$ matrix has twenty one non-zero elements. 
The eight-vertex model is related to 
spin $\frac{1}{2}$ anisotropic Heisenberg spin chain. 
Lashkevich and Pugai \cite{LaP} found that 
the correlation functions of the eight-vertex model can be obtained by 
using the free field realization of the vertex operators in the 
eight-vertex SOS model \cite{LuP}, with insertion of the nonlocal operator 
$\Lambda$, called `the tail operator'. 
In \cite{La} Lashkevich obtained integral formulae for form factors 
of the eight-vertex model. 

There are some researches which generalize the study of \cite{LaP,La}. 
The vertex operator approach for higher spin generalization of the 
eight-vertex model was presented in \cite{KKW}. 
As for higher rank generalization, the integral formulae 
for correlation functions of 
Belavin's $(\mathbb{Z}/n\mathbb{Z})$-symmetric model \cite{Bela} 
were presented 
in \cite{Bel-corr}, and those form factor formulae were presented 
in \cite{Bel-form}. 

We are interested in the form factors, originally defined as 
matrix elements of some local operators, in the 
twenty-one-vertex model. 
In this paper we will construct the tail operators 
for both diagonal and off-diagonal matrix elements 
with respect to the ground state sectors. 

Let us mention on the trigonometric limit cases of 
elliptic vertex model. In \cite{DJMO} 
the spontaneous polarization formulae of 
the higher spin analogue of the six vertex model, the trigonometric limit 
of the eight-vertex model, were obtained by using Bethe Ansatz 
method. Idzumi \cite{Idz} reproduced those formulae for spin $1$ case 
in terms of vertex operator formalism. 
In the critical limit, the spin $\frac{k}{2}$ (isotropic) 
Heisenberg spin chain 
is described by level $k$ Wess-Zumino-Witten model \cite{Reshe}, 
whose central charge is given by $c=\tfrac{3k}{k+2}$. 
Since $c=1$ for the spin $\frac{1}{2}$ case, the eight-vertex model can be 
described in terms of one boson. Spin $1$ analogue of the eight-vertex 
model (twenty-one-vertex model) 
can be described in terms of one boson and one fermion, because 
$c=\frac{3}{2}=1+\frac{1}{2}$ for $k=2$. Actually, 
Idzumi \cite{Idz}, Bougourzi and Weston \cite{BW} 
constructed level $2$ irreducible highest weight representations 
of the quantum affine Lie algebra $U_q (\widehat{\mathfrak{sl}}_2)$ 
in terms of one boson and one fermion. 

Let us turn to the elliptic case. 
Baxter \cite{8VSOS1,8VSOS2,8VSOS3} found the vertex-face transformation 
which relates the eight-vertex model and the SOS model. 
Boos et al. \cite{BJMST} 
proposed a conjectural formula for multi-point correlation functions 
of the $Z$-invariant (inhomogeneous) eight-vertex model. 
The restricted SOS (RSOS) model was constructed in \cite{ABF}. 
The higher spin generalization of RSOS model was introduced 
in \cite{FusionSOS1,FusionSOS2} on the basis of the 
fusion procedure. Kojima, Konno and Weston \cite{KKW} constructed 
vertex operator formalism for the higher spin analogue of 
the eight-vertex model, by using vertex-face transformation 
onto $k\times k$ fusion SOS model. 

The present paper was written 
in a self-contained manner so that section 2 and section 3 except 
subsection 3.5 are reviews of previous related works. In section 2 
we review the basic objects of the twenty-one-vertex model, 
the corresponding fusion face model \cite{FusionSOS1,FusionSOS2}, 
the vertex-face correspondence of these two models, and 
the tail operators which translate correlation functions and 
form factors 
of fusion SOS model into those of the twenty-one-vertex model. 
Some detail definitions 
of the models concerned are listed in Appendix A. 
In section 3 we introduce free field representations 
for $2\times 2$ fusion SOS model. The type I and the type II 
vertex operators, 
the tail operators and the CTM Hamiltonian 
can be realized in terms of bosons and fermions. 
Form factors of the twenty-one-vertex model 
can be obtained by these objects, in principle. 
Section 4 is devoted to derivation of the integral formulae 
for form factors 
of the twenty-one-vertex model. 
In order to show the relevance of our present method, we calculate the 
simplest form factor of the local operator $S^z_1$ in subsection 4.2. 
In section 5 we give some concluding remarks. 
Useful operator product expansion (OPE) formulae and 
commutation relations for basic operators 
are given in Appendix B. 
In Appendix C, the details of derivation are given for 
the free filed representations of the 
tail operators off-diagonal 
with respect to the ground state sectors.

\section{Basic objects} 

The present section aims to formulate 
the problem, thereby fixing the notation. 

\subsection{Theta functions} 

Jacobi theta function with two pseudo-periods $1$ and 
$\tau$\,(${\rm Im}\,\tau >0$) are defined as follows: 
\begin{equation}
\vartheta\left[\begin{array}{c} a \\ b \end{array} \right]
(u;\tau ): =\displaystyle\sum_{m\in \mathbb{Z}} 
\exp \left\{ \pi \sqrt{-1}(m+a)~\left[ (m+a)\tau 
+2(u+b) \right] \right\}, \label{Rieth}
\end{equation}
for $a,b\in\mathbb{R}$. 
In what follows we use the symmbols 
$\theta_1 (u; \tau ), \cdots , \theta_4 (u; \tau )$ when 
$(a,b)=(\tfrac{1}{2}, -\tfrac{1}{2}), 
(\tfrac{1}{2}, 0), (0, 0), (0, \tfrac{1}{2})$ 
on (\ref{Rieth}), respectively. 
Let $r>2$ and $\epsilon >0$ be fixed, and let 
$$
h_j^{(t)}(u):= \theta_j \left( \tfrac{u}{t}; 
\tfrac{\pi\sqrt{-1}}{\epsilon t}\right), ~~~~ (j=1,2,3,4)
$$
for $t>0$. We put $h_j^{(r)}(u)=h_j(u)$ for simplicity. 
We will use the abbreviations, 
\begin{equation}
\begin{array}{ll}
[u]:=x^{\frac{u^2}{r}-u}\Theta_{x^{2r}}(x^{2u}), ~~~~ 
& 
\{u\}:=x^{\frac{u^2}{r}-u}\Theta_{x^{2r}}(-x^{2u}), \\
{[}\!{[} u{]}\!{]}:=x^{\frac{u^2}{r}}\Theta_{x^{2r}}(x^{2u+r}), ~~~~ 
& 
\{\!\!\{ u\}\!\!\}:=x^{\frac{u^2}{r}}\Theta_{x^{2r}}(-x^{2u+r}), 
\end{array}
\end{equation}
where 
\begin{eqnarray*}
&&\Theta_{q}(z)=(z; q)_\infty 
(qz^{-1}; q)_\infty (q; q)_\infty =
\sum_{m\in\mathbb{Z}} q^{m(m-1)/2}(-z)^m, \\
&&(z; q_1 , \cdots , q_m )_\infty = 
\prod_{i_1 , \cdots , i_m \geqslant 0} 
(1-zq_1^{i_1} \cdots q_m^{i_m}). 
\end{eqnarray*}
Note that 
$$
\begin{array}{ll}
h_1 (u)
=\sqrt{\tfrac{\epsilon r}{\pi}}
\exp\,\left(-\tfrac{\epsilon r}{4}\right)[u], ~~~~ & 
h_4 (u)=\sqrt{\tfrac{\epsilon r}{\pi}}
\exp\,\left(-\tfrac{\epsilon r}{4}\right)\{u\}, \\
h_2 (u)
=\sqrt{\tfrac{\epsilon r}{\pi}}
{[}\!{[} u{]}\!{]}, ~~~~ & 
h_3 (u)=\sqrt{\tfrac{\epsilon r}{\pi}}
\{\!\!\{ u\}\!\!\},
\end{array}
$$
where $x=e^{-\epsilon}$. 

In the present paper we often use the following abbreviations: 
$$
r'=r-1, ~~ r''=r-2, ~~ [u]'=[u]\left|_{r\mapsto r-1}\right.\!, ~~ 
[u]''=[u]\left|_{r\mapsto r-2}\right., 
$$
and so on. 

\subsection{Spin $1$ analogue of the eight-vertex model}

The twenty-one-vertex model is 
constructed from the original spin $\tfrac{1}{2}$ eight-vertex model 
by fusion procedure. Let 
\begin{equation}
R(u)v_{\varepsilon_1} \otimes 
v_{\varepsilon_2}=\sum_{\varepsilon'_1, \varepsilon'_2 =\pm} 
v_{\varepsilon'_1} \otimes 
v_{\varepsilon'_2}R(u)_{\varepsilon_1\varepsilon_2}^
{\varepsilon'_1\varepsilon'_2} 
\end{equation}
be the $R$-matrix of the eight-vertex model. 
Non-zero elements of the $R$-matrix are given as follows: 
\begin{equation}
\begin{array}{ll}
R(u)_{\varepsilon\varepsilon}^{\varepsilon\varepsilon}=
\dfrac{1}{\bar{\kappa}(u)}
\dfrac{h_2^{(2r)}(1)h_2^{(2r)}(u)}{
h_2^{(2r)}(0)h_2^{(2r)}(1-u)}, 
& 
R(u)_{\varepsilon -\varepsilon}^{\varepsilon -\varepsilon} 
=-\dfrac{1}{\bar{\kappa}(u)}
\dfrac{h_2^{(2r)}(1)h_1^{(2r)}(u)}{
h_2^{(2r)}(0)h_1^{(2r)}(1-u)}, 
\\[6mm]
R(u)_{\varepsilon -\varepsilon}^{-\varepsilon \varepsilon} 
=\dfrac{1}{\bar{\kappa}(u)}
\dfrac{h_1^{(2r)}(1)h_2^{(2r)}(u)}{
h_2^{(2r)}(0)h_1^{(2r)}(1-u)}, 
& 
R(u)_{\varepsilon \varepsilon}^{-\varepsilon -\varepsilon} 
=\dfrac{1}{\bar{\kappa}(u)}
\dfrac{h_1^{(2r)}(1)h_1^{(2r)}(u)}{
h_2^{(2r)}(0)h_2^{(2r)}(1-u)}, 
\end{array}
\label{eq:abcd}
\end{equation}
where, 
\begin{eqnarray}
\bar{\kappa}(u)&=& 
\displaystyle\zeta^{-\frac{r-1}{r}}
\frac{\rho (z)}{\rho (z^{-1})}, ~~~~ 
(z=\zeta^2 =x^{2u}, x=e^{-\epsilon}) 
\label{eq:kappa} \\
\rho (z)&=&\frac{(x^2 z ; x^4 , x^{2r})_{\infty}
   (x^{2r+2}z ; x^4 , x^{2r})_{\infty}}
      {(x^4 z ; x^4 , x^{2r})_{\infty}
      (x^{2r} z ; x^4 , x^{2r})_{\infty}}. 
\nonumber 
\end{eqnarray}

Let 
\begin{equation}
R^{(1,1)}(u)v_{j_1} \otimes 
v_{j_2}=\sum_{j'_1, j'_2 =-1}^{1} 
v_{j'_1} \otimes 
v_{j'_2}R^{(1,1)}(u)_{j_1j_2}^{j'_1j'_2} 
\end{equation}
be the $R$-matrix of the twenty-one-vertex model. 
This $R^{(1,1)}(u)$ can be obtained from $R(u)$ in terms of 
fusion procedure. The following property 
\begin{equation}
PR^{(1,1)}(1)=-R^{(1,1)}(1), ~~~~ P(x\otimes y)=y\otimes x, 
\end{equation}
is important in the fusion procedure. 
The explicit expressions of the matrix elements of $R$-matrix of 
the twenty-one-vertex model are 
given 
in Appendix A. 

We assume that the parameters $u$, $\epsilon$ and $r$ 
on (\ref{eq:abcd}) and (\ref{eq:21v}) lie 
in the so-called principal regime: 
\begin{equation}
\epsilon >0, ~~ r>2, ~~ 0<u<1. 
\label{eq:principal}
\end{equation}
This is the antiferroelectric region of the parameters. 
The twenty-one-vertex model has three kinds of 
ground states labeled by $i$ for $i=0,1,2$. Accordingly, 
there are three spaces of physical states ${\cal H}^{(i)}$ ($i=0,1,2$). 
Here, the space ${\cal H}^{(i)}$ is the $\mathbb{C}$-vector space 
spanned by the half-infinite pure tensor vectors of the forms
\begin{equation}
v_{s_1}\otimes v_{s_2}\otimes 
v_{s_3}\otimes \cdots 
~~~~ \mbox{with $s_j\in \{ -1,0,1\}$, ~~ for 
$j=1,2,3,\cdots$}
\label{eq:H^i}
\end{equation}
and 
\begin{equation}
\mbox{$s_j=\left\{ \begin{array}{ll} 
1-i & (j\equiv 0 ~ \mbox{mod }2) \\ 
i-1 & (j\equiv 1 ~ \mbox{mod }2) \end{array} \right.$ for $j\gg 0$}. 
\label{eq:H^i-bc}
\end{equation}
Note that ${\cal H}^{(i)}$ is isomorphic to 
the level $2$ highest weight module of affine Lie algebra 
$A^{(1)}_1$, with the highest weight 
$$
\lambda_i :=
(2-i)\Lambda_0 +i\Lambda_1 ~~~~ (i=0,1,2), 
$$
respectively. Here, $\Lambda_i$'s ($i=0,1$) denote the fundamental weights 
of $A^{(1)}_1$. 

Let ${\cal H}^{*(i)}$ be the dual of ${\cal H}^{(i)}$ 
spanned by the half-infinite pure tensor vectors of the forms 
\begin{equation}
\cdots \otimes v_{s_{-2}}\otimes v_{s_{-1}}\otimes 
v_{s_0}
~~~~ \mbox{with $s_j\in \{ -1,0,1\}$, ~~ for 
$j=0,-1,-2,\cdots$}
\end{equation}
and 
\begin{equation}
\mbox{$s_j=\left\{ \begin{array}{ll} 
1-i & (j\equiv 0 ~ \mbox{mod }2) \\ 
i-1 & (j\equiv 1 ~ \mbox{mod }2) \end{array} \right.$ for $j\ll 0$}. 
\end{equation}

Let us consider the so-called low temperature limit 
$x\rightarrow 0$ of (\ref{eq:21v}) with $\zeta =x^{u}$ be fixed. 
Then the $R^{(1,1)}(u)$ behaves as 
\begin{equation}
R^{(1,1)}(u)^{s_1s_2}_{s'_1s'_2} 
\sim \zeta^{H(s_1, s_2)}
\delta_{s'_2}^{s_1}\delta_{s'_1}^{s_2} ~~~~  (x\rightarrow 0)
\label{eq:lowtemp-R}
\end{equation}
where 
\begin{equation}
H(s,s')=|s+s'|=\left\{ \begin{array}{ll} 
0 & (\mbox{if $(s,s')=(\pm 1,\mp 1), (0,0)$}) \\ 
1 & (\mbox{if $(s,s')=(\pm 1,0), (0,\pm 1)$}) \\ 
2 & (\mbox{if $(s,s')=(\pm 1,\pm 1)$}) \end{array} 
\right. 
\end{equation}

Thus, the South-East corner transfer matrix behaves 
\begin{equation}
A^{(i)}_{SE}(u)^{s_1s_2\cdots}
_{s'_1s'_2\cdots} \sim 
\zeta^{H^{(i)}_{CTM}(s_1, s_2, \cdots )}
\delta^{s_1}_{s'_1}
\delta^{s_2}_{s'_2}
\cdots : {\cal H}^{(i)}\rightarrow {\cal H}^{(i)}, 
\label{eq:lowtempA}
\end{equation}
in the low temperature limit $x\rightarrow 0$, where 
\begin{equation}
H^{(i)}_{CTM}(s_1 , s_2 , 
\cdots )=\sum_{j=1}^\infty jH(s_j,s_{j+1}). 
\label{eq:CTM-H}
\end{equation}
We assume that (\ref{eq:lowtempA}) is valid not only for 
low temperature limit $x\rightarrow 0$ but also for 
finite $0<x<1$\footnote{Note that the 
$u$-dependence of $R^{(1,1)}(u)$ 
is actually $\zeta$-dependence, where $\zeta=x^u$. Since 
the eigenvalues $\lambda_s$ 
of $A_{SE}^{(i)}(u)$ should be invariant under the 
shift $u\mapsto u+2\pi\sqrt{-1}/\log x$, we have 
$\lambda_s =\zeta^{n_s}$ ($n_s\in\mathbb{Z}$). 
Owing to the discreteness property of eigenvalues, (\ref{eq:lowtempA}) 
should be valid even for finite $0<x<1$, in the sense of similarity 
transformation. }. 
Likewise other three types of the corner transfer matrices 
are introduced as follows: 
\begin{equation}
\begin{array}{ll}
A^{(i)}_{NE}(u): & {\cal H}^{(i)}\rightarrow {\cal H}^{*(i)}, \\
A^{(i)}_{NW}(u): & {\cal H}^{*(i)}\rightarrow {\cal H}^{*(i)}, \\
A^{(i)}_{SW}(u): & {\cal H}^{*(i)}\rightarrow {\cal H}^{(i)}, 
\end{array}
\end{equation}
where NE, NW and SW stand for the corners North-East, North-West 
and South-West. 
It seems to be rather general \cite{ESM} that the product of four CTMs 
in the infinite lattice limit is independent of $u$: 
\begin{equation}
\rho^{(i)}=A^{(i)}_{SE}(u)A^{(i)}_{SW}(u)
A^{(i)}_{NW}(u)A^{(i)}_{NE}(u)
=x^{2H^{(i)}_{CTM}}. 
\label{eq:rho^i}
\end{equation}
The trace of $\rho^{(i)}$ coincides with the principally specialized 
character of $\lambda_i$, up to some factors \cite{DJKMO1}: 
\begin{equation}
\chi^{(i)}:={\rm tr}_{{\cal H}^{(i)}}\,(x^{2H^{(i)}_{CTM}}) =
x^i \chi_{\lambda_i} (x)=\left\{ \begin{array}{ll}
(-x^2; x^2)_\infty (-x^4; x^4)_\infty & (i=0,2) \\
(-x^2; x^2)_\infty (-x^2; x^4)_\infty & (i=1) \end{array}
\right. 
\label{eq:ch-H}
\end{equation}

Introduce the type I vertex operator by the following 
half-infinite transfer matrix 
\begin{equation}
\Phi^j (u_1 -u_2)=
\unitlength 0.5mm
\begin{picture}(100,20)
\put(15,-18){\begin{picture}(100,0)
\put(60,20){\vector(-1,0){60}}
\put(10,30){\vector(0,-1){20}}
\put(20,30){\vector(0,-1){20}}
\put(30,30){\vector(0,-1){20}}
\put(40,30){\vector(0,-1){20}}
\put(4,22){$j$}
\put(-7,18){$u_1$}
\put(9,5){$u_2$}
\put(19,5){$u_2$}
\put(29,5){$u_2$}
\put(39,5){$u_2$}
\put(45,22){$\cdots$}
\end{picture}
}
\end{picture}
\label{eq:B-I}
\end{equation}

~

\noindent Then the operator (\ref{eq:B-I}) is an intertwiner 
from ${\cal H}^{(i)}$ to ${\cal H}^{(2-i)}$. 
The type I vertex operators satisfy the following 
commutation relation: 
\begin{equation}
\Phi^{j_1} (u_1)\Phi^{j_2} (u_2)=
\sum_{j'_1,j'_2} R^{(1,1)}(u_1-u_2)^{j_1j_2}_{j'_1j'_2} 
\Phi^{j'_2} (u_2)\Phi^{j'_1} (u_1). 
\label{eq:RPhiPhi}
\end{equation}

When we consider an operator related to `creation-annihilation' process, 
we need another type of vertex operators, the type II vertex operators 
that satisfy the following commutation relations: 
\begin{equation}
\Psi^*_{j_2} (u_2)\Psi^*_{j_1} (u_1)=
\sum_{j'_1,j_2'} \Psi^*_{j'_1} (u_1)\Psi^*_{j'_2} (u_2)
S^{(1,1)}(u_1-u_2)_{j_1j_2}^{j'_1j'_2}, 
\label{eq:R'PsiPsi}
\end{equation}
\begin{equation}
\Phi^{j_1} (u_1)\Psi^*_{j_2} (u_2)=-
\Psi^*_{j_2} (u_2)\Phi^{j_1} (u_1), 
\label{eq:chiPsiPhi}
\end{equation}
where 
\begin{equation}
S^{(1,1)}(u)=
R^{(1,1)}(u)|_{r\mapsto r-2}. 
\label{eq:SXYZ}
\end{equation}

Furthermore, the type I vertex operator $\Phi^j (u)$, 
the type II vertex operator $\Psi^*_j (v)$ 
and $\rho^{(i)}$ introduced on (\ref{eq:rho^i}) satisfy 
the homogeneity relations
\begin{equation}
\Phi^{j\,(2-i,i)} (u) \rho^{(i)} =\rho^{(2-i)}\Phi^{j\,(2-i,i)} (u-2), ~~~~ 
\Psi^{*\,(2-i,i)}_j (u) \rho^{(i)} =\rho^{(2-i)}\Psi^{*\,(2-i,i)}_j (u-2), 
\label{eq:homo}
\end{equation}
and the normalization conditions 
\begin{equation}
\sum_{j=-1}^1 \Phi^{*\,(i,2-i)}_j (u) 
\Phi^{j\,(2-i,i)} (u)  =1, ~~~~ 
\Psi^{j'\,(i,2-i)} (u') \Psi^{*\,(2-i,i)}_j (u) =
\dfrac{\delta^{j'}_j}{1-x^{-2}z/z'} +O(1). 
\label{eq:normalization}
\end{equation}
Here, $z=x^{2u}$, $z'=x^{2u'}$ and 
\begin{equation}
\lambda\Phi^{*\,(2-i,i)}_j (u) =\Phi^{-j\,(2-i,i)} (u-1), ~~~~ 
\lambda^*\Psi^{*\,(2-i,i)}_j (u) =\Psi^{-j\,(2-i,i)} (u-1), 
\label{eq:dual-VO21}
\end{equation}
and $\lambda$ and $\lambda^*$ are appropriate constants 
defined later.

\subsection{$2\times 2$ fusion SOS model}

The SOS model was introduced by Baxter \cite{8VSOS1,8VSOS2,8VSOS3} 
in order to solve the eight-vertex model. 
The state variables of the SOS model take integer values. 
A pair $(a,b) \in \mathbb{Z}^2$ 
is called admissible if $b=a\pm 1$. 
Let $(a,b)$ be the state variables at adjacent sites. 
Then the pair $(a,b)$ is admissible. 
For $(a, b, c, d)\in \mathbb{Z}^{4}$ let 
$\displaystyle W 
\left[ \left. \begin{array}{cc} 
c & d \\ b & a \end{array} 
\right| u \right] $ 
be the Boltzmann weight of the SOS model for the state configuration 
$\displaystyle 
\left[ \begin{array}{cc} 
c & d \\ b & a \end{array} \right] $ 
round a face. 
Here the four states $a, b, c$ and $d$ are 
ordered clockwise from the SE corner. 
In this model $W 
\left[ \left. \begin{array}{cc} 
c & d \\ b & a \end{array} \right| 
u \right] =0~~$ 
unless the four pairs $(a,b), (a,d), (b,c)$ 
and $(d,c)$ are admissible. 
Non-zero Boltzmann weights are given as follows: 
\begin{equation}
\begin{array}{rcl}
\displaystyle W\left[ \left. \begin{array}{cc} 
k\pm 2 & k\pm 1 \\ k\pm 1 & k \end{array} \right| u \right]
& = & \dfrac{1}{\bar{\kappa }(u)}, \\[6mm]
\displaystyle W\left[ \left. \begin{array}{cc} 
k & k\pm 1 \\ k\pm 1 & k \end{array} \right| u \right]
& = & \displaystyle \dfrac{1}{\bar{\kappa }(u)}
\frac{[1][k \pm u]}{[1-u][k]}, \\[6mm]
\displaystyle W\left[ \left. \begin{array}{cc} 
k & k\pm 1 \\ k\mp 1 & k \end{array} \right| u \right]
& = & -\displaystyle \dfrac{1}{\bar{\kappa }(u)}
\frac{[u][k\pm 1]}{[1-u][k]}. 
\end{array}
\label{eq:HY}
\end{equation}

The twenty-one-vertex model 
can be transformed into $2\times 2$ fusion SOS model 
in terms of vertex-face correspondence. 
Let $(a,b)$ be the state variables of $2\times 2$ fusion SOS model 
at adjacent sites. 
Then $b=a\pm 2$, or $b=a$. 
In what follows we denote $b\sim a$ when $b-a\in \{ -2,0,2\}$. 
Non-zero Boltzmann weights 
$W_{22}(u)$ are given in Appendix A. 

Here we again assume that the parameters $u$, $\epsilon$ and $r$ 
on (\ref{eq:HY}) and (\ref{eq:BW}) lie in (\ref{eq:principal}). 
This region of the parameters is called regime III 
in the SOS-type model. 
For $k, l\in\mathbb{Z}$ and $i=0,1,2$, 
let ${\cal H}^{(i)}_{l,k}$ be the space of admissible paths 
$(k_0 , k_1, k_2, \cdots )$ such that 
\begin{equation}
k_0 =k, ~~~ k_{j+1} \sim k_{j} ~ \mbox{ for $j=0, 1, 2, \cdots$, }
\end{equation}
and 
\begin{equation}
k_j=\left\{ \begin{array}{ll} 
l+i & \mbox{($j\equiv 0$ mod $2$)} \\ 
l+2-i & \mbox{($j\equiv 1$ mod $2$)} \end{array} \right. 
\mbox{ for $j\gg 0$}. 
\label{eq:SOS-bc}
\end{equation}
Also, let ${\cal H}^{*(i)}_{l,k}$ be the space of admissible paths 
$(\cdots , k_{-2} , k_{-1}, k_{0})$ such that 
\begin{equation}
k_0 =k, ~~~ k_{j-1} \sim k_{j} ~ 
\mbox{ for $j=0, -1, -2, \cdots$, }
\end{equation}
and 
\begin{equation}
k_j=\left\{ \begin{array}{ll} 
l+i & \mbox{($j\equiv 0$ mod $2$)} \\ 
l+2-i & \mbox{($j\equiv 1$ mod $2$)} \end{array} \right. 
\mbox{ for $j\ll 0$}. 
\end{equation}

After gauge transformation \cite{FusionSOS1,FusionSOS2}, 
the Boltzmann weights $W_{22}(u)$ in 
the so-called low temperature limit 
$x\rightarrow 0$ behave as 
\begin{equation}
W_{22}\left[  \left. \begin{array}{cc} c & d \\
b & a \end{array} \right| u \right] \sim 
\delta_{bd} \zeta^{\frac{1}{2}|c-a|}. 
\label{eq:lowtemp-W}
\end{equation}
Here we take the limit $x\rightarrow 0$ 
with $\zeta =x^{u}$ be fixed. 
Let $A^{(i)}_{l.k}$, $B^{(i)}_{l.k}$, $C^{(i)}_{l.k}$ 
and $D^{(i)}_{l.k}$ be the SE, SW, NW, NE corner transfer matrix. 
Then the SE corner transfer matrix 
behaves as follows: 
\begin{equation}
A^{(i)}_{l,k}(u)_{k'_0 k'1 k'_2\cdots}^{k_0 k_1 k_2 \cdots} 
\sim \zeta^{H^{(i)}_{l,k}}
\delta_{k'_0}^{k_0}\delta_{k'_1}^{k_1}\delta_{k'_2}^{k_2}\cdots , 
{\cal H}^{(i)}_{l,k}\rightarrow {\cal H}^{(i)}_{l,k}, 
\label{eq:lowtempA_lk}
\end{equation}
in the low temperature limit $x\rightarrow 0$, where 
\begin{equation}
H^{(i)}_{l,k}(k_0 , k_1 , k_2, 
\cdots )=\dfrac{1}{2}\sum_{j=1}^\infty j|k_{j+1} -k_{j-1}|. 
\end{equation}

Likewise other three types of the corner transfer matrices 
are introduced as follows: 
\begin{equation}
\begin{array}{ll}
B^{(i)}_{l,k}(u): & {\cal H}^{*(i)}\rightarrow {\cal H}^{(i)}, \\
C^{(i)}_{l,k}(u): & {\cal H}^{*(i)}\rightarrow {\cal H}^{*(i)}, \\
D^{(i)}_{l,k}(u): & {\cal H}^{(i)}\rightarrow {\cal H}^{*(i)}. 
\end{array}
\end{equation}
It seems to be rather general \cite{ESM} that the product of four CTMs 
in the infinite lattice limit is independent of $u$: 
\begin{equation}
\rho^{(i)}_{l,k}=A^{(i)}_{l,k}(u)B^{(i)}_{l,k}(u)
C^{(i)}_{l,k}(u)D^{(i)}_{l,k}(u)
=[k]x^{2H^{(i)}_{l,k}}. 
\label{eq:r_lk}
\end{equation}

Let $k\equiv l+i+2j$ (mod $4$), where $i=0,1,2$ and $j=0,1$. 
Then the trace of $\rho^{(i)}_{l,k}$ can be given as follows 
\cite{DJKMO2}: 
\begin{equation}
\chi^{(i)}_{l,k}:={\rm tr}_{{\cal H}^{(i)}_{l,k}}\,(\rho^{(i)}_{l,k}) =
[k] c^{\lambda_i}_{\lambda_{i+2j}}(x^4) 
x^{\tfrac{r}{2r''}l^2-lk+\tfrac{r''}{2r}k^2}. 
\label{eq:ch-H_lk}
\end{equation}
Here $c^{\lambda_i}_{\lambda_j}(x^4)$ is the string function 
\cite{KP}, up to some factors. We change the definitions 
for the present purpose as follows: 
\begin{equation}
\begin{array}{rcl}
c^{\lambda_0}_{\lambda_0}(x^4)\pm c^{\lambda_0}_{\lambda_2}(x^4)
&=&\dfrac{(\mp x^2; x^4)_{\infty}}{(x^4; x^4)_\infty}, \\
c^{\lambda_1}_{\lambda_1}(x^4)=c^{\lambda_1}_{\lambda_3}(x^4)
&=&\dfrac{x^{\frac{1}{2}}(-x^4; x^4)_\infty}{(x^4; x^4)_\infty}, \\
c^{\lambda_i}_{\lambda_j}(x^4)&=&0 ~~ \mbox{(for $j\not\equiv i$ mod $2$)}. 
\end{array}
\label{eq:string-fn}
\end{equation}
Besides (\ref{eq:string-fn}) we have the following symmetry
\begin{equation}
c^{\lambda_i}_{\lambda_j}(x^4)=c^{\lambda_i}_{\lambda_{j+4}}(x^4)
=c^{\lambda_{2-i}}_{\lambda_{2-j}}(x^4). 
\label{eq:string-sym}
\end{equation}
{}From (\ref{eq:ch-H_lk}), (\ref{eq:string-fn}), 
(\ref{eq:string-sym}) and (\ref{eq:ch-H}), 
we obtain the following relation \cite{KKW}: 
\begin{equation}
\sum_{k\in l+i+2\mathbb{Z}} \chi^{(i)}_{l,k} 
=[l]'' \chi^{(i)}. 
\label{eq:chi-chi-rel}
\end{equation}

Introduce the type I vertex operator by the following 
half-infinite transfer matrix 
\begin{equation}
\hspace{2cm} 
\unitlength 1mm
\begin{picture}(100,15)
\put(0,-55){\begin{picture}(101,0)
\put(-14,59){$\Phi(u_1 -u_2)_k^{k'}=$}
\put(20,55){\vector(0,1){10}}
\put(30,55){\vector(0,1){10}}
\put(30,55){\vector(-1,0){10}}
\put(30,65){\vector(-1,0){10}}
\put(40,55){\vector(0,1){10}}
\put(40,55){\vector(-1,0){10}}
\put(40,65){\vector(-1,0){10}}
\put(50,55){\vector(0,1){10}}
\put(50,55){\vector(-1,0){10}}
\put(50,65){\vector(-1,0){10}}
\put(60,55){\vector(-1,0){10}}
\put(60,65){\vector(-1,0){10}}
\put(19,52){$k$}
\put(18,65){$k'$}
\multiput(18,60)(2,0){23}{\line(1,0){1}}
\put(17,60){\vector(-1,0){1.5}}
\multiput(25,53)(0,2){8}{\line(0,1){1}}
\multiput(35,53)(0,2){8}{\line(0,1){1}}
\multiput(45,53)(0,2){8}{\line(0,1){1}}
\put(25,52){\vector(0,-1){1.5}}
\put(35,52){\vector(0,-1){1.5}}
\put(45,52){\vector(0,-1){1.5}}
\put(24,48){$u_2$}
\put(34,48){$u_2$}
\put(44,48){$u_2$}
\put(12,59.5){$u_1$}
\end{picture}
}
\end{picture}
\label{eq:F-I}
\end{equation}

\vspace{7mm}

\noindent Then the operator (\ref{eq:F-I}) is an intertwiner 
from ${\cal H}^{(i)}_{l,k}$ to 
${\cal H}^{(2-i)}_{l,k'}$. 
The type I vertex operators satisfy the following 
commutation relation: 
\begin{equation}
\Phi (u_1)^c_b\Phi (u_2)^b_a=
\sum_{d} W_{22}\left[ \left. \begin{array}{cc} 
c & d \\ 
b & a \end{array} \right| u_1-u_2 \right]
\Phi (u_2)^{c}_d\Phi (u_1)^{d}_a . 
\label{eq:Wphiphi}
\end{equation}
The free field realization of $\Phi (u)^b_a$ was constructed 
in \cite{KKW}. See section 3.2. 

The type II vertex operators should satisfy 
the following commutation relations: 
\begin{equation}
\Psi^* (u_2)^{c}_{d}\Psi^* (u_1)^{d}_{a}=
\sum_{b} 
\Psi^* (u_1)^{c}_{b}\Psi^* (u_2)^{b}_{a} 
W''_{22}\left[ \left. \begin{array}{cc} 
c & d \\ 
b & a \end{array} \right| u_1-u_2 \right], 
\label{eq:W'psipsi}
\end{equation}
\begin{equation}
\Phi (u_1)^{k'}_k\Psi^* (u_2)^{l'}_{l}=
-\Psi^* (u_2)^{l'}_{l}\Phi (u_1)^{k'}_k , 
\label{eq:Wchipsiphi}
\end{equation}
where 
\begin{equation}
W''_{22}\left[ \left. 
\begin{array}{cc} c & d \\ b & a \end{array} \right| 
v \right]=W_{22}\left[ \left. 
\begin{array}{cc} c & d \\ b & a \end{array} \right| 
v \right] \left. \makebox{\rule[-4mm]{0pt}{11mm}} 
\right|_{r\mapsto r-2}. 
\label{eq:W''SOS}
\end{equation}

Furthermore, the type I vertex operator $\Phi (u)_k^{k'}$, 
the type II vertex operator $\Psi^* (u)_l^{l'}$ 
and $\rho^{(i)}_{l,k}$ introduced on (\ref{eq:r_lk}) satisfy 
the homogeneity relations 
\begin{equation}
\Phi^{(2-i,i)} (u)^{k'}_k \dfrac{\rho^{(i)}_{l,k}}{[k]} 
=\dfrac{\rho^{(2-i)}_{l,k'}}{[k']}\Phi^{(2-i,i)} (u-2)^{k'}_k, ~~~~ 
\Psi^{*\,(2-i,i)} (u)^{l'}_{l} \rho^{(i)}_{k,l} =
\rho^{(2-i)}_{k,l'}\Psi^{*\,(2-i,i)} (u-2)^{l'}_{l}, 
\label{eq:Whomo}
\end{equation}
and the normalization conditions 
\begin{equation}
\sum_{k'\sim k} \Phi^{*\,(i,2-i)} (u)^k_{k'} 
\Phi^{(2-i,i)} (u)^{k'}_k  =1, ~~~~ 
\Psi^{(i,2-i)} (u')^{l''}_{l'} 
\Psi^{*\,(2-i,i)} (u)^{l'}_l =\dfrac{\delta_l^{l''}}{1-x^{-2}z/z'} 
+O(1). 
\label{eq:normalizationSOS}
\end{equation}
Here, $z=x^{2u}$, $z'=x^{2u'}$ and the dual vertex operators 
$\Phi^{*} (u)^k_{k'}$ and $\Psi^{*} (u)^{l'}_l$ will be 
defined in section 3.

\subsection{Vertex-face correspondence}

Baxter \cite{8VSOS1,8VSOS2,8VSOS3} introduced the intertwining vectors 
which relate the eight-vertex model and the SOS model. 
Let 
\begin{equation}
\begin{array}{rcl}
\tau (u)^k_{k\pm 1}&=&
\displaystyle\sum_{\varepsilon=\pm} v_\varepsilon 
\tau^\varepsilon (u)^k_{k\pm 1}=\dfrac{f(u)}{\sqrt{2}} 
\begin{bmatrix} h_3^{(2r)}(k\mp u) \\ h_4^{(2r)}(k\mp u)
\end{bmatrix}, 
\end{array}
\label{eq:int-vec}
\end{equation}
where the scalar function $f(u)$ satisfies 
$$
h_1 (u)f(u)f(u-1)=1. 
$$
Then the following relation holds: (cf. Figure 1) 
\begin{equation}
R(u_1-u_2)\tau (u_1)_a^d\otimes \tau (u_2)_d^c=
\sum_{b} \tau (u_1)_b^c \otimes \tau (u_2)_a^b 
W\left[ \left. 
\begin{array}{cc} c & d \\ b & a \end{array} \right| 
u_1 -u_2 \right]. 
\label{eq:Rtt=Wtt}
\end{equation}

\unitlength 1mm
\begin{picture}(100,20)
\put(23,0){
\begin{picture}(101,0)
\put(20,3){\begin{picture}(101,0)
\put(10,10){\vector(-1,0){10}}
\put(10,0){\vector(0,1){10}}
\put(8.8,5){\vector(-1,0){10}}
\put(8.,4.4){\scriptsize{$<$}}
\put(-4.5,4.2){$u_1$}
\put(4.15,8.5){\scriptsize{$\vee$}}
\put(5,8.8){\vector(0,-1){10}}
\put(3.9,-3.8){$u_2$}
\put(-2.5,10.5){$c$}
\put(10.5,-1.5){$a$}
\put(10.5,10.1){$d$}
\put(17,4){$=\;\displaystyle\sum_{b}$} 
\end{picture}
}
\put(58,3){\begin{picture}(101,0)
\put(10,0){\vector(-1,0){10}}
\put(10,0){\vector(0,1){10}}
\put(0,0){\vector(0,1){10}}
\put(10,10){\vector(-1,0){10}}
\put(-2.,4.4){\scriptsize{$<$}}
\multiput(0,5)(2.2,0){6}{\line(1,0){1.2}}
\put(-1.2,5){\vector(-1,0){2.5}}
\put(-7,4.2){$u_1$}
\put(4.15,-1.5){\scriptsize{$\vee$}}
\multiput(5,0)(0,2.2){6}{\line(0,1){1.2}}
\put(5,-1.2){\vector(0,-1){2.5}}
\put(3.9,-5.8){$u_2$}
\put(10.5,10.1){$d$}
\put(-2.5,10.5){$c$}
\put(10.5,-1.5){$a$}
\put(-2.5,-1.8){$b$}
\end{picture}
}
\end{picture}
}
\end{picture}

\vspace{2mm}

\begin{center}
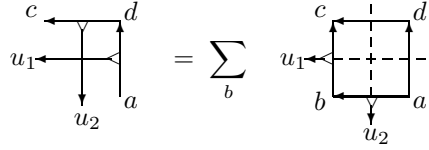
Figure 1. Picture representation of vertex-face 
correspondence. 
\end{center}

Note that the present intertwining vectors are different 
from the ones used in \cite{8VSOS1,8VSOS2,8VSOS3}, which relate 
the $R$-matrix of eight-vertex model in 
the disordered phase and Boltzmann weights $W$ of the 
SOS model in the regime III. 

Let us introduce the dual intertwining vectors (see Figure 2) 
satisfying 
\begin{equation}
\sum_{\varepsilon =\pm} \tau_\varepsilon^*  (u)^{k'}_{k}
\tau^\varepsilon (u)^{k}_{k''} =\delta_{k''}^{k'}, ~~~~ 
\sum_{k'=k\pm 1} \tau^\varepsilon (u)^{k}_{k'} 
\tau_{\varepsilon'}^* (u)^{k'}_{k} =
\delta^\varepsilon_{\varepsilon'}. \label{eq:dual-t}
\end{equation}

\unitlength 1mm
\begin{picture}(100,20)
\put(40,3){\begin{picture}(101,0)
\put(-10,4){$\displaystyle\sum_{\varepsilon =\pm}$}
\put(10.2,-3.){$k'$}
\put(-2.7,-2.8){$k$}
\put(4.15,0.1){\scriptsize{$\wedge$}}
\put(10,0){\vector(-1,0){10}}
\multiput(0,0)(0,2.2){5}{\line(0,1){1.2}}
\put(10.2,10.5){$k''$}
\put(-2.7,10.5){$k$}
\put(10,10){\vector(-1,0){10}}
\put(6,4){$\varepsilon$}
\put(4.15,8.6){\scriptsize{$\vee$}}
\put(5,1.2){\line(0,1){7.6}}
\put(5,0){\line(0,-1){1}}
\put(5,-1.5){\vector(0,-1){2}}
\put(4.22,-6){$u$}
\put(15.,4.){$=\delta^{k''}_{k'}$,}
\end{picture}
}
\put(88,3){\begin{picture}(101,0)
\put(-13,4){$\displaystyle\sum_{k'=k\pm 1}$}
\put(-3,4){$k$}
\put(11,4){$k'$}
\put(10,5){\vector(-1,0){10}}
\put(4.15,5.1){\scriptsize{$\wedge$}}
\put(4.15,3.6){\scriptsize{$\vee$}}
\put(5,6.4){\line(0,1){3.6}}
\put(6,7){$\varepsilon'$}
\put(5,3.6){\vector(0,-1){3.6}}
\put(6,2){$\varepsilon$}
\put(4.2,-2.2){$u$}
\put(17,4.){$=\delta^{\varepsilon'}_{\varepsilon}$.}
\end{picture}
}
\end{picture}

\vspace{3mm}

\begin{center}
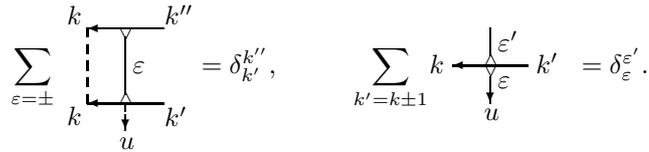
Figure 2. Picture representation of the dual intertwining 
vectors. 
\end{center}

{}From (\ref{eq:Rtt=Wtt}) and (\ref{eq:dual-t}), we have (cf. Figure 3) 
\begin{equation}
\tau^*(u_{1})^{b}_{c}\otimes \tau^*(u_{2})^{a}_{b}
R(u_{1}-u_2 )=
\displaystyle\sum_{d} 
W\left[ \left. \begin{array}{cc} 
c & d \\ b & a \end{array} \right| u_{1}-u_2 \right]
\tau^*(u_{1} )^{a}_{d}\otimes \tau^*(u_{2} )^{d}_{c}. 
\label{eq:dJMO}
\end{equation}

\unitlength 1mm
\begin{picture}(100,20)
\put(23,0){
\begin{picture}(101,0)
\put(20,3){\begin{picture}(101,0)
\put(10,0){\vector(-1,0){10}}
\put(0,0){\vector(0,1){10}}
\put(5,6.4){\line(0,1){3.6}}
\put(-.2,4.4){\scriptsize{$>$}}
\put(1.4,5){\line(1,0){10}}
\put(0,5){\line(-1,0){1}}
\put(-1.5,5){\vector(-1,0){2}}
\put(-6.5,4.2){$u_1$}
\put(4.15,.1){\scriptsize{$\wedge$}}
\put(5,1.4){\line(0,1){10}}
\put(5,0){\line(0,-1){1}}
\put(5,-1.5){\vector(0,-1){2}}
\put(4,-5.8){$u_2$}
\put(-2.5,10.5){$c$}
\put(10.5,-1.5){$a$}
\put(-2.5,-1.8){$b$}
\put(17,4){$=\;\displaystyle\sum_{d}$} 
\end{picture}
}
\put(56,3){\begin{picture}(101,0)
\put(10,0){\vector(-1,0){10}}
\put(10,0){\vector(0,1){10}}
\put(0,0){\vector(0,1){10}}
\put(10,10){\vector(-1,0){10}}
\put(9.8,4.4){\scriptsize{$>$}}
\multiput(10,5)(-2.2,0){6}{\line(-1,0){1.2}}
\put(11.6,5){\line(1,0){2}}
\put(-1.2,5){\vector(-1,0){2}}
\put(-7,4.2){$u_1$}
\put(4.15,10.1){\scriptsize{$\wedge$}}
\put(5,11.5){\line(0,1){2}}
\multiput(5,10)(0,-2.2){6}{\line(0,-1){1.2}}
\put(5,-1.2){\vector(0,-1){2}}
\put(4,-5.3){$u_2$}
\put(10.5,10.1){$d$}
\put(-2.5,10.5){$c$}
\put(10.5,-1.5){$a$}
\put(-2.5,-1.8){$b$}
\end{picture}
}
\end{picture}
}
\end{picture}

\vspace{3mm}

\begin{center}
Figure 3. Vertex-face correspondence by dual intertwining 
vectors. 
\end{center}

Intertwining vectors which relate the twenty-one-vertex model 
and the $2\times 2$ fusion SOS model can be 
constructed by fusion procedure. In what follows 
let us denote these vectors for the fusion models 
by $t(u)^{k'}_{k}$ and $t^*(u)^{k'}_{k}$. The explicit expressions 
of these fused intertwining vectors are given in Appendix A. 

Let 
\begin{equation}
t''{}^* (u)^{l'}_{l}:=t^* (u; \epsilon , r-2)^{l'}_{l}. 
\label{eq:t'*}
\end{equation}
Then we have 
\begin{equation}
\displaystyle R^{(1,1)}(u_1 -u_2 ) t(u_1 )^{d}_{a}\otimes 
t(u_2 )^{c}_{d}
=\displaystyle\sum_{b} 
t(u_1 )_{b}^{c}\otimes t(u_2 )_{a}^{b} 
W_{22}\left[ \left. \begin{array}{cc} 
c & d \\ b & a \end{array} \right| u_1 -u_2 \right]. 
\label{eq:sJMO(R)}
\end{equation}
and 
\begin{equation}
\displaystyle t''{}^*(u_1 )^{b}_{c}\otimes 
t''{}^*(u_2 )^{a}_{b}S^{(1,1)}(u_1 -u_2 ) 
=\displaystyle\sum_{d} 
W''_{22}\left[ \left. \begin{array}{cc} 
c & d \\ b & a \end{array} \right| u_1 -u_2 \right]
t''{}^*(u_1 )_{d}^{a}\otimes t''{}^*(u_2 )_{c}^{d}. 
\label{eq:sJMO}
\end{equation}

Let us introduce the following symbol
\begin{equation}
L\left[  \left. \begin{array}{cc} a'_0 & a'_1 \\
a_0 & a_1 \end{array} \right| u_0 \right] :=
\sum_{j=-1}^{1} t^*_j (-u_0)_{a_0}^{a_1} 
t^j (-u_0)^{a'_0}_{a'_1}. 
\label{eq:Lop}
\end{equation}
Then form (\ref{eq:inv-1}) 
\begin{equation}
L\left[  \left. \begin{array}{cc} a_0 & a'_1 \\
a_0 & a_1 \end{array} \right| u_0 \right] =
\delta_{a_1}^{a'_1}. 
\label{eq:Lop-inv}
\end{equation}
The explicit expressions of $L$ are given in Appendix A. 

Assume that $0<\Re (u_0)<2$. Then 
it follows from (\ref{eq:int-vec-1}) and (\ref{eq:dual-int-vec-1}) 
that for $i=0,1,2$, 
\begin{equation}
|t^*_{i-1}(-u_0)_{l+i}^{l+2-i}t^{i-1}(-u_0)^{l+i}_{l+2-i}|\sim 1 ~~~~ 
(x\rightarrow 0)
\end{equation}
is much greater than other products 
$t^*_{j-1}(-u_0)_{l+i}^{l+2-i}t^{j-1}(-u_0)^{l+i}_{l+2-i}$ ($j\neq i$), 
in the low temperature limit. Thus, the boundary condition 
${\cal H}^{(i)}$ of the twenty-one-vertex model 
(\ref{eq:H^i-bc}) corresponds to that of ${\cal H}^{(i)}_{l,k}$ 
of the $2\times 2$ fusion SOS model (\ref{eq:SOS-bc}).

\subsection{Tail operators and commutation relations}

Tail operators were originally introduced in \cite{LaP,La}, 
in order to translate correlation functions of the eight-vertex model 
into those of SOS model. Tail operators for higher spin case 
were constructed in \cite{KKW}, and those for higher rank case 
were constructed in \cite{Bel-corr,Bel-form}. 

Let us introduce the intertwining operators between 
${\cal H}^{(i)}$ and ${\cal H}^{(i)}_{l,k}$: 
\begin{equation}
\begin{array}{rcl}
T(u_0){}^{lk}&=&
\displaystyle\prod_{j=1}^\infty 
t^{s_j}(-u_0){}^{k_{j-1}}_{k_{j}}: 
{\cal H}^{(i)}\rightarrow {\cal H}^{(i)}_{l,k}, \\
T(u_0 ){}_{lk}&=&
\displaystyle\prod_{j=1}^\infty 
t^*_{s_j}(-u_0 ){}_{k_{j-1}}^{k_{j}}: 
{\cal H}^{(i)}_{l,k}\rightarrow {\cal H}^{(i)}. 
\end{array}
\label{eq:T^_}
\end{equation}
{}From (\ref{eq:Rtt=Wtt1}) and (\ref{eq:dSOS_2}) 
the following intertwining relations hold: 
\begin{equation}
T(u_0)^{lk'} \Phi^j (u) =\sum_k t^j 
(u-u_0 ){}_{k}^{k'}\Phi (u)^{k'}_kT(u)^{lk}, 
\label{eq:T^Phi}
\end{equation}
\begin{equation}
T(u_0)_{lk'} \Phi (u)^{k'}_k =\sum_{j=-1}^1 t^*_j 
(u-u_0 ){}^{k}_{k'}\Phi^j (u)T(u_0)_{lk}. 
\label{eq:T_phi}
\end{equation}

Tail operator is defined by the product of these two objects 
(see Figure 4): 
\begin{equation}
\Lambda (u_0)_{lk}^{l'k'}=T(u_0)^{l' k'}T(u_0)_{lk}: 
{\cal H}^{(i)}_{l,k}\rightarrow {\cal H}^{(i)}_{l'k'}. 
\label{eq:L=TT'}
\end{equation}

\unitlength 1.4mm
\begin{picture}(100,20)
\put(-18,0){\begin{picture}(101,0)
\put(18,10){$\Lambda(u_0){}_{lk}^{l'k'}=$}
\multiput(65,5)(-10,0){3}{\vector(-1,0){10}}
\multiput(66,5)(2,0){5}{\line(1,0){1}}
\multiput(65,15)(-10,0){3}{\vector(-1,0){10}}
\multiput(66,15)(2,0){5}{\line(1,0){1}}
\put(106,5){\vector(-1,0){10}}
\put(106,15){\vector(-1,0){10}}
\put(96,5){\vector(-1,0){10}}
\put(96,15){\vector(-1,0){10}}
\multiput(107,5)(2,0){2}{\line(1,0){1}}
\multiput(107,15)(2,0){2}{\line(1,0){1}}
\put(86,5){\vector(-1,0){10}}
\put(86,15){\vector(-1,0){10}}
\put(34,2.3){$k$}
\put(44,2.3){$k_1$}
\put(54,2.3){$k_2$}
\put(64,2.3){$k_3$}
\put(34,16.){$k'$}
\put(44,16.2){$k'_1$}
\put(54,16.2){$k'_2$}
\put(64,16.2){$k'_3$}
\put(75,2.3){$l_1$}
\put(85,2.3){$l_2$}
\put(95,2.3){$l_1$}
\put(105,2.3){$l_2$}
\put(75,16.){$l'_1$}
\put(85,16){$l'_2$}
\put(95,16.){$l'_1$}
\put(105,16){$l'_2$}
\put(39.4165,5.){\scriptsize{$\wedge$}}
\put(39.4165,13.9){\scriptsize{$\vee$}}
\put(40,6.0){\line(0,1){7.9}}
\put(40,5){\line(0,-1){1}}
\put(40,3.5){\vector(0,-1){2}}
\put(38.,-.3){$-u_0$}
\put(49.4165,5.){\scriptsize{$\wedge$}}
\put(49.4165,13.9){\scriptsize{$\vee$}}
\put(50,6.){\line(0,1){7.9}}
\put(59.4165,5.){\scriptsize{$\wedge$}}
\put(59.4165,13.9){\scriptsize{$\vee$}}
\put(60,6.){\line(0,1){7.9}}
\put(80.4165,5.){\scriptsize{$\wedge$}}
\put(80.4165,13.9){\scriptsize{$\vee$}}
\put(81,6.){\line(0,1){7.9}}
\put(90.4165,5.){\scriptsize{$\wedge$}}
\put(90.4165,13.9){\scriptsize{$\vee$}}
\put(91,6.){\line(0,1){7.9}}
\put(100.4165,5.){\scriptsize{$\wedge$}}
\put(100.4165,13.9){\scriptsize{$\vee$}}
\put(101,6.){\line(0,1){7.9}}
\end{picture}
}
\end{picture}

\vspace{3mm}

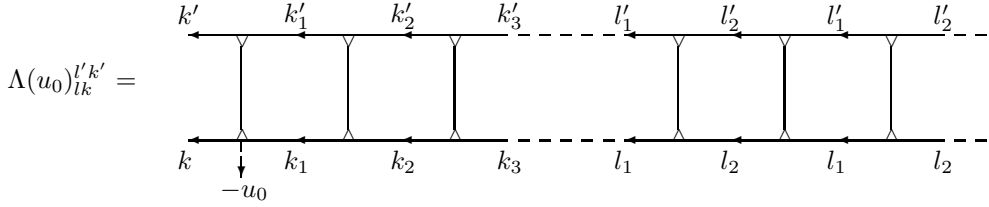
Figure 4. Tail operator $\Lambda(u_0 ){}^{l'k'}_{lk}$. 
The upper (resp. lower) half stands for $T(u_0 ){}^{l' k'}$ (resp. 
$T(u_0 ){}_{lk}$). 

Here, $l_1=l+i$, $l_2=l+2-i$, $l'_1=l'+i$ and $l'_2=l'+2-i$. 

\vspace{7mm}

{}From (\ref{eq:T^Phi}), (\ref{eq:T_phi}) and (\ref{eq:L=TT'}), 
we have 
\begin{equation}
\Lambda (u_0 )^{l'c}_{lb}\Phi (u)^b_a=
\sum_{d\sim c}L\left[ \left. \begin{array}{cc} c & d \\
b & a \end{array} \right| u_0-u \right]
\Phi (u)^c_d\Lambda (u_0)^{l'd}_{la}. 
\label{eq:Lambda-phi}
\end{equation}

Furthermore, consider the algebra 
\begin{equation}
\Psi^*_j (u)T(u_0)_{lk'}  =\sum_{l'\sim l} 
T(u_0)_{l'k}\Psi^*(u)_l^{l'}t^{*''}_j 
(u-u_0 -\Delta u_0){}^{l}_{l'}, 
\label{eq:T_psi}
\end{equation}
\begin{equation}
\Psi^* (u)^{l'}_lT(u_0)^{lk} =\sum_k T(u)^{l'k} \Psi^*_j (u) 
t^{''j} (u-u_0 -\Delta u_0){}_{l}^{l'}. 
\label{eq:T^Psi}
\end{equation}
This algebra is consistent with (\ref{eq:T^Phi}--\ref{eq:T_phi}) 
for any value of $\Delta u_0$. The value of $\Delta u_0$ will be fixed 
in the next section. {}From (\ref{eq:T_psi}), (\ref{eq:T^Psi}) 
and (\ref{eq:L=TT'}), we have 
\begin{equation}
\Psi^* (u)^{l''}_{l'}\Lambda (u_0)^{l'\,k'}_{l\,k}=
\sum_{l_1}L''\left[ \left. \begin{array}{cc} l'' & l' \\
l_1 & l \end{array} \right| u_0+\Delta u-u \right]
\Lambda (u_0)^{l''\,k'}_{l_1\,k}\Psi^* (u)^{l_1}_{l}. 
\label{eq:Lambda-psi}
\end{equation}

In what follows we suppress $l$-dependence to denote 
$\Lambda (u_0){}^{lk'}_{lk}$ by $\Lambda (u_0){}^{k'}_{k}$. 
{}From (\ref{eq:T^_}), 
(\ref{eq:L=TT'}) and (\ref{eq:Lop}) we have 
\begin{equation}
\Lambda(u_0 ){}_{lk}^{l'k'}=
\prod_{j=0}^\infty L\left[  \left. \begin{array}{cc} 
k'_j & k'_{j+1} \\
k_j & k_{j+1} \end{array} \right| u_0 \right]. 
\label{eq:Lambda}
\end{equation}
\vspace{5mm}
It is obvious from (\ref{eq:Lop-inv}), we have 
\begin{equation}
\Lambda (u_0 )^{l'k}_{lk} =\delta_l^{l'}. 
\label{eq:Lambda=1}
\end{equation}

The relation (\ref{eq:chi-chi-rel}) implies that 
\begin{equation}
\mbox{tr}_{{\cal H}^{(i)}}\, (\rho^{(i)}) 
=\dfrac{1}{[l]''}\sum_{k\in l+i+2\mathbb{Z}} 
\mbox{tr}_{{\cal H}^{(i)}_{l,k}}\, (\rho^{(i)}_{l,k}). 
\label{eq:rho-rel1}
\end{equation}
Insert unity (\ref{eq:Lambda=1}) into the RHS of (\ref{eq:rho-rel1}). 
Then we have 
\begin{equation}
\begin{array}{rcl}
\mbox{tr}_{{\cal H}^{(i)}}\, (\rho^{(i)}) 
&=&\displaystyle\sum_{k\in l+i+2\mathbb{Z}} 
\mbox{tr}_{{\cal H}^{(i)}_{l,k}}\, 
\left(\dfrac{\rho^{(i)}_{l,k}}{[l]''}T(u_0)^{l k}T(u)_{lk}\right) \\
&=&\displaystyle\sum_{k\in l+i+2\mathbb{Z}} 
\mbox{tr}_{{\cal H}^{(i)}}\, 
\left(T(u)_{lk}\dfrac{\rho^{(i)}_{l,k}}{[l]''}T(u_0)^{l k}\right). 
\end{array}
\label{eq:rho-rho-rel}
\end{equation}
Thus in what follows we assume that 
\begin{equation}
\rho^{(i)}=\sum_{k\in l+i+2\mathbb{Z}} 
T (u)_{lk} \dfrac{\rho^{(i)}_{l,k}}{[l]''}  T(u)^{lk}. 
\label{eq:rho-rel}
\end{equation}

\section{Free filed realization}

One of the most standard ways to calculate correlation functions 
and form factors 
is the vertex operator approach \cite{JMbk} 
on the basis of free field representation. 
The face type elliptic quantum group 
${\cal B}_{q,\lambda}(\widehat{\mathfrak{sl}}_2)$ was 
introduced in \cite{JKOS1}. 
The elliptic algebra $U_{q,p}(\widehat{\mathfrak{sl}}_2)$ 
associated with fusion SOS models was defined in \cite{Ko}, 
and its free field representations were constructed in \cite{Ko,JKOS2}. 
Using these representations we derive the free field representation 
of the tail operator in this section. 

\subsection{Bosons and fermions}

Let us consider the bosons
$\beta_m\,(m \in \mathbb{Z}\backslash \{0\})$ 
with the commutation relations
\begin{equation}
[\beta_m,\beta_{m'}]
=m\dfrac{[r''m]_x}{[rm]_x} \delta_{m+m',0}. 
\label{eq:comm-B}
\end{equation}
Here the symbol $[a]_x$ stands for
$(x^a-x^{-a})/(x-x^{-1})$.
The relation between the present $\beta_m$ and the 
bosons $a_m$ in \cite{KKW} is as follows: 
\begin{equation}
\beta_m =\left\{ \begin{array}{ll} 
\dfrac{m[r''m]_x}{[2m]_x [rm]_x} a_m & (m>0) \\
\dfrac{mx^{-2m}}{[2m]_x} a_m & (m<0) \end{array}
\right. 
\end{equation}

We will deal with the bosonic Fock spaces 
${\cal{F}}^{(i)}_{l,k}, (l,k \in \mathbb{Z})$
generated by $\beta_{-m} (m>0)$ and $e^{\alpha}$, $e^{\beta}$
over the vacuum vectors $|l,k\rangle$ :
\begin{eqnarray*}
{\cal{F}}^{(i)}_{l,k}=
\mathbb{C}[\beta_{-1}, \beta_{-2}, \cdots ]\otimes 
\left( \oplus_{n\in \mathbb{Z}} \mathbb{C} e^{\lambda_i+n\alpha+m\beta} 
\right) |l,k\rangle,
\end{eqnarray*}
where
\begin{eqnarray*}
\beta_m|l,k\rangle&=&0 ~(m>0),\\
e^{\pm\alpha}|l,k\rangle &=& | l,k\pm 2\rangle , \\
e^{\pm\beta}|l,k\rangle &=& | l\pm 2,k\rangle .
\end{eqnarray*}
Let $K$ and $L$ be the operators which act 
diagonally on ${\cal{F}}^{(i)}_{l,k}$: 
$$
K|l,k\rangle =k|l,k\rangle , ~~~~ 
L|l,k\rangle =l|l,k\rangle . 
$$

Furthermore, let us consider the fermions 
\begin{equation} 
\phi (w) =\sum_{m} \phi_m w^{-m}
\end{equation}
with the anticommutation relations 
\begin{equation}
[\phi_m, \phi_{m'}]_+ =
\delta_{m+m',0} \dfrac{x^{2m}+x^{-2m}}{x+x^{-1}}. 
\end{equation} 
We refer to $\phi_m$'s for $m\in\mathbb{Z}+\tfrac{1}{2}$ as 
Neveu-Schwarz fermions, and $\phi_m$'s for $m\in\mathbb{Z}$ as 
Ramond fermions. 
Let 
$$
{\cal F}^{\phi}=\left\{ 
\begin{array}{ll}
\mathbb{C}[\phi_{-\frac{1}{2}}, \phi_{-\frac{3}{2}}, \cdots ] & 
(\mbox{for $i=0,2$}) \\ 
\mathbb{C}[\phi_{-1}, \phi_{-2}, \cdots ] & (\mbox{for $i=1$}) 
\end{array} \right. 
$$
be the fermionic Fock space. 

Note that the following anticommutation relation holds: 
\begin{equation}
[\phi (w_1), \phi (w_2)]_+ =\dfrac{1}{x+x^{-1}}\left( 
\delta \left( \frac{x^2w_2}{w_1} \right)+
\delta \left( \frac{x^2w_1}{w_2} \right) 
\right). 
\label{eq:phi-phi-anti}
\end{equation}
Here we use $\phi_0^2 =1/(x+x^{-1})$ for Ramond fermion sector. 

The total space of states ${\cal H}^{(i)}_{l,k}$ is isomorphic to 
\begin{equation}
{\cal H}^{(i)}_{l,k}={\cal F}^{(i)}_{l,k}\otimes 
{\cal F}^{\phi}. 
\label{eq:total-sp}
\end{equation}

\subsection{Free field realization of type I vertex operators}

Let us introduce the following basic operators
\begin{equation}
\begin{array}{rcl}
\Phi_1 (u) &=&
z^{\frac{r''}{2r}}:\exp \left( -\displaystyle\sum_{m\neq 0} 
\frac{\beta_m}{m}z^{-m}\right):e^\alpha 
z^{-\frac{1}{2}L+\frac{r''}{2r}K}, \\ 
A(v)&=&w^{\frac{r''}{2r}}:\exp \left( 
\displaystyle\sum_{m\neq 0} 
\frac{\beta_m}{m}w^{-m}\right):e^{-\alpha} 
w^{\frac{1}{2}L-\frac{r''}{2r}K}\phi (w), 
\end{array}
\end{equation}
where $z=x^{2u}$, $w=x^{2v}$. 
As for some useful OPE formulae and commutation relations, 
see Appendix B. 

Then the type I vertex operators (half transfer matrices) 
on ${\cal{H}}^{(i)}_{l,k}$ can be realized in terms of bosons and 
fermions: 
\begin{equation}
\begin{array}{rcl}
\Phi(u)_k^{k+2} &=& 
\dfrac{[1]}{[k][k+1]} \Phi_1 (u), \\[3mm]
\Phi(u)_k^k &=& \dfrac{[2]}{[k-1][k+1]} \Phi_1 (u) X(u), \\[3mm]
\Phi(u)_k^{k-2} &=& \dfrac{[1]}{[k][k-1]} \Phi_1 (u) 
X(u)^2, 
\end{array}
\label{eq:type-I}
\end{equation}
where 
\begin{equation}
X(u)=\oint_C 
\dfrac{dw}{2\pi \sqrt{-1} w} A(v) 
\dfrac{[v-u-K]}{[v-u-1]}
\label{eq:X-def}
\end{equation}
Considering the denominators 
$[v-u-1]$'s together with the OPE 
formulae (\ref{eq:PhiA-OPE}), 
the expressions (\ref{eq:type-I}) has poles at 
$w=x^{\pm (2+2nr)}z\,(n \in \mathbb{Z}_{\geqslant 0})$. 
The integral contour $C$ for $w$-integration is 
the anti-clockwise circle 
such that all integral variables lie in the 
common convergence domain; i.e., the contour $C$ 
encircles the poles at $w=x^{2+2nr}z\,(n 
\in \mathbb{Z}_{\geqslant 0})$, but not the poles at 
$w=x^{-2-2nr}z\,(n \in \mathbb{Z}_{\geqslant 0})$. 

Let 
\begin{equation}
Y(u)=-\oint_{C} 
\dfrac{dw}{2\pi \sqrt{-1} w} A(v) 
\dfrac{[v-u+2-K]}{[v-u+1]}. 
\label{eq:Y-def}
\end{equation}
Then we can rewrite (\ref{eq:type-I}) as follows: 
\begin{equation}
\begin{array}{rcl}
\Phi(u)_k^{k+2} &=& 
\dfrac{[1]}{[k][k+1]} \Phi_1 (u), \\[3mm]
\Phi(u)_k^k &=& \dfrac{[2]}{[k-1][k+1]} Y(u) \Phi_1 (u), \\[3mm]
\Phi(u)_k^{k-2} &=& \dfrac{[1]}{[k][k-1]} Y(u)^2 \Phi_1 (u). 
\end{array}
\label{eq:type-I'}
\end{equation}

Note that 
\begin{equation}
\Phi(u)_k^{k'}: {\cal{H}}^{(i)}_{l,k} 
\longrightarrow {\cal{H}}^{(2-i)}_{l,k'}. 
\label{eq:k-shift}
\end{equation}
These type I vertex operators satisfy 
the following commutation relations on 
${\cal{H}}^{(i)}_{l,k}$: 
\begin{eqnarray}
\Phi(u_1)_b^c \Phi(u_2)_a^b
=\sum_{d}W_{22}\left[\left.
\begin{array}{cc}
c & d \\
b & a
\end{array}\right|u_1-u_2 \right]
{\Phi}(u_2)_d^c{\Phi}(u_1)_a^d. 
\label{eq:CR-I}
\end{eqnarray}

Dual vertex operators are likewise defined as follows: 
\begin{equation}
\begin{array}{rcl}
\Phi^*(u)_{k-2}^{k}&=& \dfrac{1}{\lambda}\Phi_1 (u-1), \\[3mm] 
\Phi^*(u)_{k}^{k}&=& 
\dfrac{1}{\lambda}\Phi_1 (u-1)X(u-1), \\[3mm] 
\Phi^*(u)_{k+2}^{k}&=& \dfrac{1}{\lambda} 
\Phi_1 (u-1)X(u-1)^2. 
\end{array}
\label{eq:type-I*}
\end{equation}
Here the normalization factor can be determined as 
$$
\lambda =\dfrac{(x^{2r'};x^{2r})_\infty^2}{(x+x^{-1})
(x^2; x^{2r})^2_\infty (x^{2r};x^{2r})_\infty^{3}}, 
$$ 
such that $\Phi (u)_k^{k'}$ and $\Phi^* (u)^k_{k'}$ satisfy the 
inversion relation: 
\begin{equation}
\sum_{k'\sim k} \Phi^* (u)^k_{k'} 
\Phi (u)_k^{k'}=1. 
\label{eq:dualrel}
\end{equation}
As explained below (\ref{eq:X-def}), the integral contour 
$C=C_u$ actually depends on $u$. On eqs. (\ref{eq:type-I*}) 
the $w$-integration contour $C_{u-1}$ of $X(u-1)$ 
encircles the poles at $w=x^{2nr}z\,(n 
\in \mathbb{Z}_{\geqslant 0})$, but not the poles at 
$w=x^{-4-2nr}z\,(n \in 
\mathbb{Z}_{\geqslant 0})$. 
Note that 
\begin{equation}
\Phi^* (u)^k_{k'}: {\cal{H}}^{(i)}_{l,k'} 
\longrightarrow {\cal{H}}^{(2-i)}_{l,k}. 
\label{eq:k-shift*}
\end{equation}
A level $2$ representation of the elliptic algebra 
$U_{x,p}(\widehat{\mathfrak{sl}}_2)$ was obtained 
in terms of one free boson and one free fermion in \cite{Czech,Annecy}. 

\subsection{Free field realization of type II vertex operators}

Let us introduce the following basic operators
\begin{eqnarray}
\Psi^*_1 (u) &=&
z^{\frac{r}{2r''}}:\exp \left( \displaystyle\sum_{m\neq 0} 
\dfrac{[rm]_x}{[r''m]_x}\frac{\beta_m}{m}z^{-m}\right):e^\beta 
z^{\frac{r}{2r''}L-\frac{1}{2}K}, \\ 
B(v)&=&w^{\frac{r}{2r''}}:\exp \left( -
\displaystyle\sum_{m\neq 0} 
\dfrac{[rm]_x}{[r''m]_x}\frac{\beta_m}{m}w^{-m}\right):e^{-\beta} 
w^{-\frac{r}{2r''}L+\frac{1}{2}K}\phi (w), \nonumber
\end{eqnarray}
\begin{equation}
\begin{array}{rcl}
X^*(u)&=&\displaystyle\oint_{C'} 
\dfrac{dw}{2\pi \sqrt{-1} w} B(v) 
\dfrac{[v-u+L]''}{[v-u+1]''} \\
Y^*(u)&=&-\displaystyle\oint_{C'} 
\dfrac{dw}{2\pi \sqrt{-1} w} B(v) 
\dfrac{[v-u+L-2]''}{[v-u-1]''}
\end{array}
\label{eq:Y^*-def}
\end{equation}
The integral contour $C'$ for $X^*(u)$ 
encircles the poles at $w=x^{-2+2nr}z\,(n 
\in \mathbb{Z}_{\geqslant 0})$, but not the poles at 
$w=x^{2-2nr}z\,(n \in \mathbb{Z}_{\geqslant 0})$. 

Then the type II vertex operators 
on ${\cal{H}}^{(i)}_{l,k}$ can be realized in terms of bosons and 
fermions: 
\begin{equation}
\begin{array}{rcl}
\Psi^*(u)_l^{l+2} &=& 
\Psi^*_1 (u), \\[3mm]
\Psi^*(u)_l^l &=& \Psi^*_1 (u) X^*(u)=Y^*(u)\Psi^*_1 (u) , \\[3mm]
\Psi^*(u)_l^{l-2} &=& \Psi^*_1 (u) 
X^*(u)^2=Y^*(u)^2\Psi^*_1 (u), 
\end{array}
\label{eq:type-II}
\end{equation}
where $z=x^{2u}$, $w=x^{2v}$. 
As for some useful OPE formulae and commutation relations, 
see Appendix B. 

Note that 
\begin{equation}
\Psi^*(u)_l^{l'}: {\cal{H}}^{(i)}_{l,k} 
\longrightarrow {\cal{H}}^{(2-i)}_{l',k}. 
\label{eq:l-shift}
\end{equation}
These type II vertex operators satisfy 
the following commutation relations on 
${\cal{H}}^{(i)}_{l,k}$: 
\begin{eqnarray}
\Psi^*(u_2)_d^c \Psi^*(u_1)_a^d
=\sum_{b}W''_{22}\left[\left.
\begin{array}{cc}
c & d \\
b & a
\end{array}\right|u_1-u_2 \right]
{\Psi}^*(u_1)_b^c{\Psi}^*(u_2)_a^b. 
\label{eq:CR-II}
\end{eqnarray}

Dual vertex operators are likewise defined as follows: 
\begin{equation}
\begin{array}{rcl}
\Psi(u)_{l}^{l+2}&=& \dfrac{1}{\lambda^*}
\dfrac{[1]''}{[l]''[l+1]''}\Psi^*_1 (u-1), \\[3mm] 
\Psi(u)_{l}^{l}&=& \dfrac{1}{\lambda^*}
\dfrac{[2]''}{[l-1]''[l+1]''}\Psi^*_1 (u-1)X^*(u-1), \\[3mm] 
\Psi(u)_{l}^{l-2}&=& \dfrac{1}{\lambda^*} 
\dfrac{[1]''}{[l]''[l-1]''}\Psi^*_1 (u-1)X^*(u-1)^2. 
\end{array}
\label{eq:type-II*}
\end{equation}
Here the normalization factor can be determined as 
$$
\lambda^* =-\dfrac{(x^{2r'};x^{2r''})_\infty (x^{2r};x^{2r''})_\infty}
{(x+x^{-1})
(x^{-2}; x^{2r''})_\infty (x^{2r''};x^{2r''})_\infty^{4}}, 
$$ 
such that $\Psi (u)_l^{l'}$ and $\Psi^* (u)^l_{l'}$ satisfy the 
inversion relation: 
\begin{equation}
\Psi (u')_l^{l'}\Psi^* (u)^l_{l''} 
=\dfrac{\delta^{l'}_{l''}}{1-x^{-2}z/z'}+O(1). 
\label{eq:dualrel-II}
\end{equation}

For later convenience, we also introduce another type of 
basic operator: 
\begin{equation}
W (v)=w^{\frac{2}{rr''}}:\exp \left( -\displaystyle\sum_{m\neq 0} 
\dfrac{[2m]_x}{[r''m]_x}\frac{\beta_m}{m}w^{-m}\right):e^{-
\alpha-\beta} 
w^{\frac{K}{r}-\frac{L}{r''}}. 
\label{eq:IxII}
\end{equation}
Concerning useful OPE formulae and commutation relations, 
see Appendix B.

\subsection{Free field realization of tail operators -- diagonal sectors}

Another ingredient of the present scheme is the tail 
operators $\Lambda (u_0)_{lk}^{l'k'}$. In this paper we use 
a different normalization from the one used in \cite{KKW}. 
Thus we briefly explain how to derive free field representations 
of $\Lambda (u_0)_{lk}^{l'k'}$. 

First let $l'=l$, that is diagonal with respect to 
the ground state sectors. 
When $k'\leqslant k-2$, let us consider (\ref{eq:Lambda-phi}) 
for $(a,b,c)=(k,k+2,k')$: 
\begin{equation}
\Lambda (u_0 )^{k'}_{k+2}\Phi (u)^{k+2}_{k}=
\sum_{k''\sim k'}L\left[ \left. \begin{array}{cc} k' & k'' \\
k+2 & k \end{array} \right| u_0-u \right]
\Phi (u)^{k'}_{k''}\Lambda (u_0)^{k''}_{k}. 
\label{eq:Lambda-Phi}
\end{equation}
Here, we briefly denote $\Lambda (u_0 )^{lk'}_{lk}$ by 
$\Lambda (u_0 )^{k'}_{k}$. 
It follows from (\ref{eq:L-op-exp}) that $L(u_0-u)$ has simple 
poles at $u_0-u=\pm\frac{1}{2}$. Note that 
$$
\left. [u_0-u+\tfrac{1}{2}]L\left[ \left. \begin{array}{cc} k' & k'' \\
k & k-2 \end{array} \right| u_0-u \right]\right|_{u_0=u-\tfrac{1}{2}}
$$
for $k''=k', k'\pm 2$ are all equal. Thus if we assume that 
the LHS of (\ref{eq:Lambda-Phi}) has no pole at $u_0 =u-\tfrac{1}{2}$, 
we have the following necessary conditions: 
\begin{equation}
\sum_{k''\sim k'}
\Phi (u)^{k'}_{k''}\Lambda (u-\tfrac{1}{2})^{k''}_{k}=0, 
\label{eq:Lambda-Phi-nec}
\end{equation}
i.e., 
\begin{equation}
\dfrac{[1]\Phi_1 (u)\Lambda (u-\tfrac{1}{2})^{k'-2}_{k}}{[k'-2][k'-1]}
+\dfrac{[2]\Phi_1 (u)X(u)\Lambda (u-\tfrac{1}{2})^{k'}_{k}}{[k'+1][k'-1]}
+\dfrac{[1]\Phi_1 (u)X(u)^2
\Lambda (u-\tfrac{1}{2})^{k'+2}_{k}}{[k'+1][k'+2]}=0. 
\label{eq:Lambda-Phi-nec'}
\end{equation}

Let $k'=k-2$. Then the LHS of (\ref{eq:Lambda-Phi-nec'}) 
contains $\Lambda (u-\tfrac{1}{2})^{k}_{k}=1$. 
By changing $k'=k-2, k-4, k-6, \cdots$, we can 
solve (\ref{eq:Lambda-Phi-nec'}) iteratively as follows: 
\begin{equation}
\begin{array}{rcl}
\Lambda (u_0)^{k-2s}_{k}
&=&\displaystyle (-X(u_0+\tfrac{1}{2}))^s 
\dfrac{[s+1][k-2s][k-s+1]}{[1][k][k+1]}. 
\end{array}
\label{eq:Bose-Lambda}
\end{equation}
Here we use the identity: 
$$
\dfrac{[1][s+1][k-s+1]}{[k-2s-1]}-
\dfrac{[2][s+2][k-2s-2][k-s]}{[k-2s-1][k-2s-3]}
+\dfrac{[1][s+3][k-s-1]}{[k-2s-3]}=0. 
$$
Furthermore, we can check that 
(\ref{eq:Bose-Lambda}) for generic $u_0$ 
satisfies (\ref{eq:Lambda-Phi}). 

Eq. (\ref{eq:Bose-Lambda}) is expressions of 
$\Lambda (u_0)^{k'}_{k}$ for $k'\leqslant k$. 
When $k'>k$, we should realize another 
free field representation of ${\cal H}^{(i)}_{l,k}$ on the 
Fock space ${\cal F}^{(i)}_{-l,-k}\otimes {\cal F}^{\phi}$. Then 
$\Lambda (u_0)^{k+2s}_{k}$ can be identified with 
$\Lambda (u_0)^{-k-2s}_{-k}$, in addition to the identification 
$\Phi (u)_k^{k'}$ and $\Phi^*(u)_k^{k'}$ with 
$\Phi (u)_{-k}^{-k'}$ and $\Phi^*(u)_{-k}^{-k'}$, respectively. 
Note that the expression (\ref{eq:Bose-Lambda}) was obtained 
in \cite{KKW} for general spin $K/2$ ($K\times K$-fused) 
SOS model.

Correlation functions in the twenty-one-vertex model 
can be constructed in terms of type I vertex operators of fusion SOS model 
and tail operators as follows: 
\begin{equation}
\begin{array}{cl}
& \dfrac{1}{\chi^{(i)}} 
{\rm tr}_{{\cal H}^{(i)}} ( 
\Phi^*_{j_{1}}(u_{1})\cdots \Phi^*_{j_{n}}(u_{n})
\Phi^{j_{n}}(u_{n})\cdots \Phi^{j_{1}}(u_{1})\rho^{(i)}) \\
=& \dfrac{1}{\chi^{(i)}} \displaystyle\sum_{k\in l+i+2\mathbb{Z}} 
{\rm tr}_{{\cal H}^{(i)}_{l,k}} \left( T(u_0)^{lk}
\Phi^*_{j_{1}}(u_{1})\cdots \Phi^*_{j_{n}}(u_{n})
\Phi^{j_{n}}(u_{n})\cdots \Phi^{j_{1}}(u_{1}) T(u_0)_{lk}
\dfrac{\rho^{(l,k)}}{[l]''}\right) \\
=& \dfrac{1}{\chi^{(i)}} 
\displaystyle\sum_{k, k_1, \cdots , k_{2n}}
t^*_{j_{1}}(u_{1}-u_0)^k_{k_{2n}}
\cdots t^*_{j_{n}}(u_{n}-u_0)^{k_{n+2}}_{k_{n+1}} 
t^{j_{n}}(u_{n}-u_0)^{k_{n+1}}_{k_{n}} \cdots 
t^{j_{1}}(u_{1}-u_0)^{k_2}_{k_1} \\
\times & 
{\rm tr}_{{\cal H}^{(i)}_{l,k}}\, 
\left( \Phi^* (u_{1})^k_{k_{2n}} \cdots \Phi^* (u_{n})^{k_{n+2}}_{k_{n+1}} 
\Phi (u_{n})^{k_{n+1}}_{k_{n}} 
\cdots \Phi (u_{1})^{k_2}_{k_{1}} \Lambda (u_0)^{k_1}_{k}
\dfrac{\rho^{(l,k)}}{[l]''}\right)
\end{array}
\label{eq:CORR-vf}
\end{equation}
Here, the sum on the third line is taken over 
$$
\{ k, k_{2n}, \cdots , k_1 | k_1 \sim k_2, \cdots , 
k_{2n}\sim k; k\in l+i+2\mathbb{Z}\}, 
$$
and 
we use (\ref{eq:T^Phi}), (\ref{eq:T_phi}), (\ref{eq:rho-rel}) and 
(\ref{eq:L=TT'}). 

\subsection{Free field realization of tail operators -- off-diagonal sectors}

In this subsection let us consider 
the tail operators for $\Lambda (u_0)_{lk}^{l'k'}$ with $l'\neq l$, 
that is off-diagonal with respect to 
the ground state sectors. 

Let $k'=k$. Then we have $l'=l$ from (\ref{eq:Lambda=1}). 
Let $k'<k$ (resp. $k'>k$) and $k'\equiv k$ (mod $2$). Then we have 
\begin{equation}
\Lambda (u_0)_{lk}^{l'k'}=0, 
\end{equation}
unless $l'\leqq l$ (resp. $l'\geqq l$). 
Actually, if $\Lambda (u_0)_{lk}^{l'k'}\neq 0$ 
for e.g., $k'<k$ and $l'>l$, there must exist a number $j$ such that 
$k'_j=k_j$ and therefore $k'_m=k_m$ for $\forall m\geqq j$ which implies 
$l'=l$ from (\ref{eq:Lambda=1}). 

Let $k'<k$ and $l'=l$ on (\ref{eq:Lambda-psi}). Firstly let $l''=l+2$. 
Then eq. (\ref{eq:Lambda-psi}) reduces to 
\begin{equation}
\Psi^* (u)^{l+2}_{l}\Lambda (u_0 )^{lk'}_{lk}=
\Lambda (u_0 )^{l+2\,k'}_{l+2\,k}\Psi^* (u)^{l+2}_{l}. 
\label{eq:Lambda-Psi1}
\end{equation}
This relation holds from (\ref{eq:Bose-Lambda}). 

Secondly let $l''=l$. Then eq. (\ref{eq:Lambda-psi}) reduces to 
\begin{equation}
[\Psi^* (u)^{l}_{l}, \Lambda (u_0 )^{lk'}_{lk}]=
\Lambda (u_0 )^{l\,k'}_{l+2\,k}\Psi^* (u)^{l+2}_{l}
L''\left[ \left. \begin{array}{cc} l & l \\
l+2 & l \end{array} \right| u_0+\Delta u-u \right]. 
\label{eq:Lambda-Psi2}
\end{equation}
{}Since $\Psi^* (u)^{l}_{l}=Y(u) \Psi^* (u)^{l+2}_{l}$ and 
(\ref{eq:Lambda-Psi1}), eq. (\ref{eq:Lambda-Psi2}) implies 
\begin{equation}
[Y(u), \Lambda (u_0 )^{lk'}_{lk}]=
\Lambda (u_0 )^{l\,k'}_{l+2\,k}
\dfrac{[1]''[u_0+\Delta u-u+l+\tfrac{1}{2}]''}{[l+2]''
[u_0+\Delta u-u-\tfrac{1}{2}]''}. 
\label{eq:Lambda-Psi2'}
\end{equation}

Thus we find $\Delta u=0$ and 
\begin{equation}
\begin{array}{rcl}
\Lambda (u_0)^{l-2\,k-2s}_{lk}
&=&\displaystyle \dfrac{[l]''}{[1]''}
\dfrac{[s][s+1][k-s+1][k-s][k-2s]}{\partial [0][1]^2[k][k+1]} \\
&\times& W_-(u_0) (-X(u_0-\tfrac{1}{2}))^{s-1}. 
\end{array}
\label{eq:BoseFermi-Lambda1}
\end{equation}
Here $\partial [0]=(x^{2r}; x^{2r})_\infty^2$, 
and 
\begin{equation}
W_- (u_0) =W\left( u_0-\tfrac{r-3}{2} \right). 
\label{eq:W_-}
\end{equation}


Thirdly let $l''=l-2$. Then eq. (\ref{eq:Lambda-psi}) reduces to 
\begin{equation}
\begin{array}{rcl}
[\Psi^* (u)^{l-2}_{l}, \Lambda (u_0 )^{k'}_{k}]&=&
\Lambda (u_0 )^{l-2\,k'}_{l\,k}\Psi^* (u)^{l+2}_{l}
L''\left[ \left. \begin{array}{cc} l-2 & l \\
l & l \end{array} \right| u_0-u \right] \\[5mm]
&+&
\Lambda (u_0 )^{l-2\,k'}_{l+2\,k}\Psi^* (u)^{l+2}_{l}
L''\left[ \left. \begin{array}{cc} l-2 & l \\
l+2 & l \end{array} \right| u_0-u \right]. 
\end{array}
\label{eq:Lambda-Psi3}
\end{equation}

By solving (\ref{eq:Lambda-Psi3}) we find 
\begin{equation}
\begin{array}{rcl}
\Lambda (u_0)^{l-4\,k-2s}_{lk}
&=&\displaystyle \dfrac{[l]''}{[1]''}
\dfrac{[s][s+1][k-s+1][k-s][k-2s]}{\partial [0][1]^2[k][k+1]} \\
&\times& X^*(u_0+\tfrac{1}{2})W_-(u_0) (-X(u_0-\tfrac{1}{2}))^{s-1}. 
\end{array}
\label{eq:BoseFermi-Lambda2}
\end{equation}

In general we obtain 
\begin{equation}
\begin{array}{rcl}
\Lambda (u_0)^{l-2t\,k-2s}_{lk}
&=&\displaystyle \dfrac{[l]''}{[1]''}
\dfrac{[s][s+1][k-s+1][k-s][k-2s]}{\partial [0][1]^2[k][k+1]} \\
&\times& X^*(u_0+\tfrac{1}{2})^{t-1}W_-(u_0) (-X(u_0-\tfrac{1}{2}))^{s-1}. 
\end{array}
\label{eq:BoseFermi-Lambda}
\end{equation}
Concerning details of derivation, see Appendix C. 

Eq. (\ref{eq:BoseFermi-Lambda}) is valid 
for $k<k$ and $l'<l$. 
When $k'>k$ and $l'>l$, we should construct the free field representation of 
$\Lambda (u_0)^{l'k'}_{lk}$ on another realization of 
${\cal H}^{(i)}_{l,k}$ on the 
Fock space ${\cal F}^{(i)}_{-l,-k}\otimes {\cal F}^{\phi}$. Then 
$\Lambda (u_0)^{l+2t\,k+2s}_{l\,k}$ can be identified with 
$\Lambda (u_0)^{-l-2t\,-k-2s}_{-l\,-k}$, 
in addition to the identification 
$\Phi (u)_k^{k'}$ and $\Psi^* (u)_l^{l'}$ with 
$\Phi (u)_{-k}^{-k'}$ and $\Psi^* (u)_{-l}^{-l'}$, respectively. 

\subsection{Free field realization of CTM Hamiltonian}

We can realize the CTM Hamiltonian of $2\times 2$ fusion SOS model 
in terms free fields as follows: 

\begin{equation}
H^{(i)}_{l,k}=H^{(l,k)}_{a}+H^{(i)}_{\phi}, 
\label{eq:totalCTM}
\end{equation}
where 
\begin{equation}
\begin{array}{rcl}
\dfrac{1}{2}H^{(l,k)}_{a}&=&\displaystyle\sum_{m=1}^\infty 
\dfrac{[rm]_x}{[r''m]_x} \beta_{-m}\beta_m +\dfrac{1}{4} 
\left( \dfrac{r}{2r''}L^2-KL+\dfrac{r''}{2r}K^2 \right), \\
\dfrac{1}{2}H^{(i)}_{\phi}&=& 
\displaystyle\sum_{n>0} n\dfrac{x+x^{-1}}{x^{2n}+x^{-2n}} 
\phi_{-n}\phi_n +\dfrac{i(2-i)}{8}. 
\label{eq:Bose/FermiCTM}
\end{array}
\end{equation}

Let us examine the validity of these expressions. First of all, 
(\ref{eq:totalCTM}) satisfies the homogeneity relation 
\begin{equation}
\Phi^{(2-i,i)} (u)_k^{k'}x^{2H^{(i)}_{l,k}}=
x^{2H^{(2-i)}_{l,k'}}\Phi^{(2-i,i)} (u-2)_k^{k'}, ~~~~ 
\Psi^{*\,(2-i,i)} (u)_l^{l'}x^{2H^{(i)}_{l,k}}=
x^{2H^{(2-i)}_{l',k}}\Psi^{*\,(2-i,i)} (u-2)_l^{l'}. 
\end{equation}
Secondly, the traces on the bosonic/fermionic Fock space are 
given as follows: 
\begin{equation}
\begin{array}{rcl}
\mbox{tr}_{{\cal F}_{l,k}^{(i)}} \left( 
x^{2H^{(l,k)}_{a}} \right)
\mbox{tr}_{{\cal F}_{\phi}} \left( 
x^{2H^{(i)}_{\phi}} \right) 
&=&
x^{\tfrac{r}{2r''}l^2-kl+\tfrac{r''}{2r}k^2}
\times \left\{ \begin{array}{ll} 
c^{\lambda_i}_{\lambda_2}+
c^{\lambda_i}_{\lambda_0} & (i=0,2) \\
c^{\lambda_1}_{\lambda_1} & (i=1) \end{array} \right. 
\end{array}
\label{eq:bose/fermi-tr}
\end{equation}
which implies (\ref{eq:rho-rel1}). From these checks 
we conclude that ${\cal H}^{(i)}_{l,k}={\cal F}^{(i)}_{l,k}\otimes 
{\cal F}_{\phi}$ and $\rho^{(i)}_{l,k}=[k] x^{2H^{(i)}_{l,k}}$. 

The fermionic trace formulae are given as follows \cite{Idz}: 
\begin{equation}
\begin{array}{cl}
&F^{(i)} (w_1, w_2):=\mbox{tr}_{{\cal F}_{\phi}} \left( 
:\phi (w_1) \phi (w_2): x^{2H^{(i)}_{\phi}} \right) \\[3mm]
=&\left\{ 
\begin{array}{ll} 
\displaystyle\dfrac{(-x^2; x^4)_\infty}{x+x^{-1}}
\sum_{m>0} \left\{ \left( \dfrac{x^2w_1}{w_2} \right)^m
-
\left( \dfrac{x^2w_2}{w_1} \right)^m 
 \right\} & (i=0,2) \\[3mm]
\displaystyle\dfrac{x^{1/2}(-x^4; x^4)_\infty}{x+x^{-1}}
\sum_{m>0} \left\{ \left( \dfrac{x^2w_1}{w_2} \right)^m
-\left( \dfrac{x^2w_2}{w_1} \right)^m \right\} & (i=1)
\end{array} \right. 
\end{array}
\label{eq:fermi-tr}
\end{equation}

\section{Form factors} 

\subsection{Integral formulae}

We are now in a position to write down integral 
formulae for form factors, 
matrix elements of some local operators. 
For simplicity, let us choose $S^z_1$ at the center site 
of the lattice as a local operator: 
\begin{equation}
S^z_1=\sum_{j=-1}^j jE^{(1)}_{j j} 
\label{eq:l-op0}
\end{equation}
The free field representation 
of $S^z$ is given by 
\begin{equation}
\hat{S^z_1}=\sum_{j=-1}^j j\Phi^{*}_{j} (u)
\Phi^{j} (u). 
\label{eq:l-op}
\end{equation}
The corresponding form factors 
with $m$ `charged' particles: 
\begin{equation}
F^{(i)}_{m}(S^z_1; u_1 , \cdots , u_{m})_{
j_{1} \cdots j_{m}}=\dfrac{1}{\chi^{(i)}} \mbox{Tr}_{
{\cal H}^{(i)}}\, \left( 
\Psi^{*}_{j_{1}} (u_{1})
\cdots \Psi^{*}_{j_m} (u_m ) \hat{S^z_1} \rho^{(i)} 
\right). 
\label{eq:d-ff}
\end{equation}
Note that the local operator (\ref{eq:l-op0}) commute 
with the type II vertex operators because of 
(\ref{eq:l-op}) and (\ref{eq:PhiPsi-OPE}--\ref{eq:BA-OPE}). 

{}From the construction in Sec. 3, 
we can rewrite (\ref{eq:d-ff}) as follows: 
\begin{equation}
\begin{array}{cl}
&F^{(i)}_{m}(S^z_1; u_1 , \cdots , u_{m})_{
j_{1} \cdots j_{m}} \\
=&\displaystyle \sum_{l_1,\cdots , 
l_m} t''{}^*_{j_1} 
\left(u_1-u_0\right){}_{
l}^{l_1}\cdots t''{}^*_{j_m} 
\left(u_m-u_0\right){}_{
l_{m-1}}^{l_m} F^{(i)}_{m}(S^z_1; u_1 , \cdots , u_{m})_{
ll_{1} \cdots l_{m}}, 
\end{array}
\label{eq:ff-rep}
\end{equation}
where 
\begin{equation}
\begin{array}{cl}
&F^{(i)}_{m}(S^z_1; u_1 , \cdots , u_{m})_{
ll_{1} \cdots l_{m}} \\
=&\dfrac{1}{\chi^{(i)}}\displaystyle\sum_{
k\equiv l+i\,(2)}
\sum_{k_1 k_2} \sum_{j=-1}^1 j
t^*_{j}(u-u_0)^{k}_{k_1} t^{j}(u-u_0)^{k_{1}}_{k_2} \\
\times&\displaystyle \mbox{Tr}_{{\cal H}^{(i)}_{l,k}}\, 
\left( \Psi^* (u_1)^l_{l_1} \cdots \Psi^* (u_m)^{l_{m-1}}_{l_m}
\Phi^* (u)^{k}_{k_1} \Phi (u)^{k_{1}}_{k_2}
\Lambda (u_0)_{l\,k}^{l_m k_2} \dfrac{[k]x^{2H^{(i)}_{l,k}}}{[l]''} \right). 
\end{array}
\label{eq:ff-rep-l}
\end{equation}
Note that eq. (\ref{eq:ff-rep}) can be inverted as follows: 
\begin{equation}
\begin{array}{cl}
&F^{(i)}_{m}(S^z_1; u_1 , \cdots , u_{m})_{
ll_{1} \cdots l_{m}} \\
=&\displaystyle \sum_{j_1,\cdots , 
j_m} t''{}^{j_1} 
\left(u_1-u_0\right){}^{
l}_{l_1}\cdots t''{}^{j_m} 
\left(u_m-u_0\right){}^{
l_{m-1}}_{l_m} F^{(i)}_{m}(S^z_1; u_1 , \cdots , u_{m})_{
j_{1} \cdots j_{m}}. 
\end{array}
\label{eq:ff-rep-inv}
\end{equation}

Free filed representations of the tail operators 
$\Lambda (u)_{l\,k}^{l'\,k'}$'s 
have been constructed in section 3, besides those of 
all other operators $\Phi (u)_k^{k'}$, $\Phi^* (u)_k^{k'}$, 
$\Psi^*(u_j)_{l}^{l'}$'s and 
$H^{(i)}_{l,k}$ on (\ref{eq:ff-rep-l}) were also given in section 3. 
Integral formulae for form factors of any local operators 
can be therefore obtained for form factors 
of spin $1$ analogue of the eight-vertex model, 
in principle. 

\subsection{Calculation of two-point form factors} 

It is very difficult 
to obtain general integral formulae (\ref{eq:ff-rep}), 
as Lashkevich said in \cite{La}. 
In order to 
show the relevance of the present scheme, we calculate the 
simplest form factor of the local operator $S^z_1$ in this subsection. 

Let us consider (\ref{eq:ff-rep-l}) for $i=2$, $m=2$, $l_1=l-2$ and 
$l_2=l-4$. Since $l_2<l$, the tail operator 
$\Lambda (u_0)_{l\,k}^{l_2 k_2}$ vanishes unless 
$k_2<k$. Thus, the sum with respect to $k_1$ and $k_2$ 
should be taken over $(k_1, k_2)=(k-2, k-4)$, $(k-2, k-2)$, 
$(k, k-2)$. We notice that the form factors (\ref{eq:ff-rep}) in the 
twenty-one-vertex model should be $u_0$-independent. 
For simplicity of calculation, let $u_0\rightarrow u-\tfrac{3}{2}$. 
By taking the sum with respect to $j=\pm 1$ 
and $(k_1, k_2)$ we have 
\begin{equation}
\begin{array}{cl}
&F^{(2)}_{2}(S^z_1; u_1 , u_{2})_{
l\,l-2\,l-4} \\
=&\dfrac{1}{2\chi^{(2)}\lambda}\dfrac{\{0\} 
[2][u-u_0-\tfrac{5}{2}]}{[1]''\partial [0] 
[u-u_0-\tfrac{1}{2}][u-u_0+\tfrac{1}{2}]}
\displaystyle\sum_{
k\equiv l\,(2)}
\dfrac{\{k+u-u_0-\tfrac{3}{2}\}}{[k+1]} \\
\times&\displaystyle\oint_{C'}\dfrac{dw_2}{2\pi\sqrt{-1}w_2} 
\dfrac{[v_2-u_0-\tfrac{5}{2}+l]''}{[v_2-u_0+\tfrac{1}{2}]''} 
\oint_{C}\dfrac{dw_1}{2\pi\sqrt{-1}w_1}
\dfrac{[v_1-u-k][v_1-u_0+\tfrac{1}{2}]}{[v_1-u][v_1-u_0-\tfrac{5}{2}]} \\
\times&\displaystyle 
T(u_1, u_2, u, u_0, v_1, v_2), 
\end{array}
\label{eq:ff-rep-l/j,k_1,k_2summed}
\end{equation}
where $T(u_1, u_2, u, u_0, v_1, v_2)$ is a trace function 
\begin{equation}
T(u_1, u_2, u, u_0, v_1, v_2)=
\mbox{Tr}_{{\cal H}^{(2)}_{l,k}}\, 
\left( \Psi^*_1(u_1) \Psi^*_1 (u_2)
\Phi_1 (u-1) \Phi_1 (u) A(v_1) B(v_2) W_-(u_0) x^{2H^{(2)}_{l,k}} \right). 
\label{eq:tr-fn}
\end{equation}
Here, the integral contour $C$ encircles the poles at 
$x^{2rn}z$ and $x^{1+2rn}z_0$ ($n\geqslant 0$) but not 
$x^{-2-2rn}z$ nor $x^{5-2r(n+1)}z_0$ ($n\geqslant 0$); 
the integral contour $C'$ encircles the poles at 
$x^{-1+2r''n}z_0$ ($n\geqslant 0$). 

{}From the expression of the fermionic trace (\ref{eq:fermi-tr}) 
the integral with respect to $w_1$ can be performed as follows: 
\begin{equation}
\begin{array}{cl}
&F^{(2)}_{2}(S^z_1; u_1 , u_{2})_{
l\,l-2\,l-4} \\
=&\dfrac{1}{2\chi^{(2)}\lambda}\dfrac{\{0\} 
[2][u-u_0-\tfrac{5}{2}]}{[1]''\partial [0] 
[u-u_0-\tfrac{1}{2}][u-u_0+\tfrac{1}{2}]}
\displaystyle\sum_{
k\equiv l\,(2)}
\dfrac{\{k+u-u_0-\tfrac{3}{2}\}}{[k+1]} \\
\times&\displaystyle\left( 
\oint_{x^{-r''}C'}\dfrac{dw_2}{2\pi\sqrt{-1}w_2} -
\oint_{x^{r''}C'}\dfrac{dw_2}{2\pi\sqrt{-1}w_2} \right) G(v_2) \\
\times&\displaystyle 
\mbox{Tr}_{{\cal F}^{(2)}_{l,k}}\, 
\left( \Psi^*_1(u_1) \Psi^*_1 (u_2)
\Phi_1 (u-1) \Phi_1 (u) W(v_2) W_-(u_0) 
x^{2H^{(l,k)}_{a}} \right), 
\end{array}
\label{eq:ff-rep-l/w_1-integrated}
\end{equation}
where 
$$
G(v_2)=\dfrac{(-x^2; x^4)_\infty}{x^{-2}-x^2}
\dfrac{[v_2-u_0+\tfrac{r-7}{2}+l]''}{[v_2-u_0+\tfrac{r-1}{2}]''} 
\dfrac{[v_2-u+\tfrac{r}{2}-k][v_2-u_0+\tfrac{r+1}{2}]}{
[v_2-u+\tfrac{r}{2}][v_2-u_0+\tfrac{r-5}{2}]}. 
$$
Thus, the difference of the two integrals with respect to $w_2$ 
on (\ref{eq:ff-rep-l/w_1-integrated}) 
can be evaluated by the residue at $w_2= x^{-r}z$ and 
$w_2=x^{1-r}z_0$\footnote{Note that 
the contour $x^{-r''}C'$ does not encircle the point 
$w_2=x^{5-r}z_0$. }. The former residue vanishes because of 
$\Phi_1(u-1) W(v_2)=0$ at $v_2=u-\tfrac{r}{2}$. Hence 
we have 
\begin{equation}
\begin{array}{cl}
&F^{(2)}_{2}(S^z_1; u_1 , u_{2})_{
l\,l-2\,l-4} \\
=&\dfrac{1}{2\chi^{(2)}\lambda}\dfrac{\{0\} 
[u-u_0-\tfrac{5}{2}][l-3]''}{[1]''\partial [0] 
\partial [0]'' [u-u_0-\tfrac{1}{2}][u-u_0+\tfrac{1}{2}]}
\displaystyle\sum_{
k\equiv l\,(2)} \{k+u-u_0-\tfrac{3}{2}\} \\
\times&\displaystyle \dfrac{(-x^2; x^4)_\infty}{x^{-2}-x^2}
\mbox{Tr}_{{\cal F}^{(2)}_{l,k}}\, 
\left( \Psi^*_1(u_1) \Psi^*_1 (u_2)
\Phi_1 (u-1) \Phi_1 (u) W_-(u_0-1) W_-(u_0) 
x^{2H^{(l,k)}_{a}} \right), 
\end{array}
\label{eq:ff-rep-l/w_2-integrated}
\end{equation}
where $\partial [0]''=
(x^{2r''}; x^{2r''})_\infty$. 

By using OPE formulae in Appendix B and the method of 
trace calculation explained in \cite{JMbk}, we obtain 
\begin{equation}
\begin{array}{cl}
&F^{(2)}_{2}(S^z_1; u_1 , u_{2})_{
l\,l-2\,l-4}= 
c x^{-7
-\tfrac{11}{r''}} 
z_1^{-\tfrac{r}{2r''}}z_2^{-\tfrac{3r}{2r''}}z^{\tfrac{2r}{r''}} \\
\times&[l-3]'' \displaystyle\sum_{
k\equiv l\,(2)} \{k\}
x^{(u_1+u_2-2u_0)(\tfrac{r}{r''}l-k)} 
x^{\tfrac{rl^2}{2r''}
-kl+\tfrac{r''k^2}{2r}} \\
\times&(x^{-2}z_2/z_1; x^{2r''})_\infty 
(x^{2r'}z_1/z_2; x^{2r''})_\infty 
(x^{2}z_2/z_1; x^{4})_\infty (x^{2}z_1/z_2; x^{4})_\infty \\
\times&\displaystyle\prod_{j=1}^2 
\dfrac{f^* (u_j-u_0)}{(x^{-2}z/z_j; x^{2})_\infty (x^4z_j/z; x^{2})_\infty}, 
\end{array}
\label{eq:sumint}
\end{equation}
where 
$$
c=\dfrac{(x^2; x^2)_\infty^2 (x^2; x^4)_\infty^2}
{(-x^4; x^4)_\infty^2 (x^4; x^4)_\infty}\dfrac{
(-x^{2r}; x^{2r})_\infty^2}{(x^{2r}; x^{2r})_\infty^2} 
\dfrac{(x^{2r'}; x^{2r''})_\infty^3}{
(x^{2r''-2}; x^{2r''})_\infty (x^{2r''}; x^{2r''})_\infty^2}, 
$$
and 
\begin{equation}
f^* (u)=\dfrac{1}{(x^{-1}z^{-1}; x^{2r''})_\infty 
(x^{2r-3}z; x^{2r''})_\infty (x^{-3}z^{-1}; x^{2r''})_\infty 
(x^{2r-1}z; x^{2r''})_\infty}. 
\label{eq:df-f^*}
\end{equation}

By substituting 
$$
\begin{array}{cl}
&\displaystyle\sum_{
k\equiv l\,(2)} \{k\}
x^{(u_1+u_2-2u_0)(\tfrac{r}{r''}l-k)} 
x^{\tfrac{rl^2}{2r''}
-kl+\tfrac{r''k^2}{2r}} \\
=&x^{-\tfrac{1}{r''}(u_1+u_2-2u_0)^2+u_1+u_2-2u_0} 
\{ l+u_1+u_2-2u_0 \}'' \{ 2u_0-u_1-u_2 \}_2. 
\end{array}
$$
into (\ref{eq:sumint}), we get 
\begin{equation}
\begin{array}{cl}
& F^{(2)}_{2}(S^z_1; u_1 , u_{2})_{
l\,l-2\,l-4} =\dfrac{\pi x^{-r''/2}}{2\epsilon r''}
c(u_1, u_2, u) \{ u_1+u_2-2u_0 \}_2 \\
\times& \left \{ h_1^{(2r'')} (2l+u_1+u_2-2u_0 -3) 
h_2^{(2r'')} (u_1+u_2-2u_0+3) \right. \\ 
-& \left. h_2^{(2r'')} (2l+u_1+u_2-2u_0-3) 
h_1^{(2r'')} (u_1+u_2-2u_0+3) \right\}. 
\end{array}
\label{eq:summed-integrated}
\end{equation}
where 
$$
\begin{array}{cl}
& c(u_1, u_2, u) = cx^{-7
-\tfrac{11}{r''}} 
z_1^{-\tfrac{r}{2r''}}z_2^{-\tfrac{3r}{2r''}}z^{\tfrac{2r}{r''}} \\
\times&(x^{-2}z_2/z_1; x^{2r''})_\infty 
(x^{2r'}z_1/z_2; x^{2r''})_\infty 
(x^{2}z_2/z_1; x^{4})_\infty (x^{2}z_1/z_2; x^{4})_\infty \\
\times&x^{-\tfrac{1}{r''}(u_1+u_2-2u_0)^2+u_1+u_2-2u_0} 
\displaystyle\prod_{j=1}^2 
\dfrac{f^* (u_j-u_0)}{(x^{-2}z/z_j; x^{2})_\infty (x^4z_j/z; x^{2})_\infty}, 
\end{array}
$$
Note that 
\begin{equation}
\begin{array}{cl}
&t''{}^{1}(u_1-u_0)^l_{l-2} t''{}^{-1}(u_2-u_0)_{l-4}^{l-2} 
-t''{}^{-1}(u_1-u_0)^l_{l-2} t''{}^{1}(u_2-u_0)_{l-4}^{l-2} \\ 
=& \dfrac{h_1^{(2r'')} (2l+u_1+u_2-2u_0-3) 
h_1^{(2r'')} (u_2-u_1-2)
h_2^{(2r'')} (0) h_2^{(2r'')} (2)}{4h_1''(u_1-u_0+\tfrac{1}{2})
h_1''(u_2-u_0+\tfrac{1}{2})h_1''(u_1-u_0+\tfrac{3}{2})
h_1''(u_2-u_0+\tfrac{3}{2})}, \\
&t''{}^{1}(u_1-u_0)^l_{l-2} t''{}^{1}(u_2-u_0)_{l-4}^{l-2} 
-t''{}^{-1}(u_1-u_0)^l_{l-2} t''{}^{-1}(u_2-u_0)_{l-4}^{l-2} \\ 
=& \dfrac{h_2^{(2r'')} (2l+u_1+u_2-2u_0-3) 
h_2^{(2r'')} (u_2-u_1-2)
h_2^{(2r'')} (0) h_2^{(2r'')} (2)}{4h_1''(u_1-u_0+\tfrac{1}{2})
h_1''(u_2-u_0+\tfrac{1}{2})h_1''(u_1-u_0+\tfrac{3}{2})
h_1''(u_2-u_0+\tfrac{3}{2})}, 
\end{array}
\label{eq:t''-addition}
\end{equation}
where $h_j''(u):=h_j(u)|_{r\mapsto r-2}$ ($j=1,2,3,4$). 
Thus, eq. (\ref{eq:summed-integrated}) can be reduced as follows: 
\begin{equation}
\begin{array}{cl}
& F^{(2)}_{2}(S^z_1; u_1 , u_{2})_{
l\,l-2\,l-4} \\
=& \dfrac{[u_1-u_0+\tfrac{1}{2}]''
[u_2-u_0+\tfrac{1}{2}]''[u_1-u_0+\tfrac{3}{2}]''
[u_2-u_0+\tfrac{3}{2}]''}{x^{-r''/2}{[}\!{[} 1{]}\!{]}
\{\!\!\{ 1\}\!\!\}}
c(u_1, u_2, u) \{ u_1+u_2-2u_0 \}_2 \\
\times& \left\{ \left( t''{}^{1}(u_1-u_0)^l_{l-2} 
t''{}^{-1}(u_2-u_0)_{l-4}^{l-2} 
-t''{}^{-1}(u_1-u_0)^l_{l-2} t''{}^{1}(u_2-u_0)_{l-4}^{l-2} \right) 
\mbox{\rule[-1mm]{0pt}{6mm}} \right. \\
\times & \dfrac{h_2^{(2r'')} (u_1+u_2-2u_0+3)}{
h_1^{(2r'')} (u_2-u_1-2)} -
\dfrac{h_1^{(2r'')} (u_1+u_2-2u_0+3)}{
h_2^{(2r'')} (u_2-u_1-2)} \\ 
\times & \left. \mbox{\rule[-1mm]{0pt}{6mm}} 
\left( t''{}^{1}(u_1-u_0)^l_{l-2} 
t''{}^{1}(u_2-u_0)_{l-4}^{l-2} 
-t''{}^{-1}(u_1-u_0)^l_{l-2} t''{}^{-1}(u_2-u_0)_{l-4}^{l-2} \right) 
\right\}. 
\end{array}
\label{eq:l-addition}
\end{equation}

By comparing (\ref{eq:l-addition}) and (\ref{eq:ff-rep-inv}), 
we obtain 
\begin{equation}
\begin{array}{rcl}
F^{(2)}_{2}(S^z_1; u_1 , u_{2})_{
\pm 1,\mp 1} 
&=& \pm
d(u_1, u_2, u) \{ u_1+u_2-2u+3 \}_2
\dfrac{h_2^{(2r'')} (u_1+u_2-2u+6)}{
h_1^{(2r'')} (u_2-u_1-2)}, \\ 
F^{(2)}_{2}(S^z_1; u_1 , u_{2})_{
\pm 1,\pm 1} 
&=& \mp
d(u_1, u_2, u) \{ u_1+u_2-2u+3 \}_2
\dfrac{h_1^{(2r'')} (u_1+u_2-2u+6)}{
h_2^{(2r'')} (u_2-u_1-2)}, \\ 
\end{array}
\label{eq:ff-final}
\end{equation}
where 
$$
\begin{array}{cl}
& d(u_1, u_2, u)=c\dfrac{(x^{2r''}; x^{2r''})_\infty^2}{
(x^{2r''+2}; x^{4r''})_\infty (x^{2r''-2}; x^{4r''})_\infty} \\
\times & x^{\tfrac{1}{r''}(u_1-u_2)^2} x^{
\tfrac{r''}{2}+6+\tfrac{4}{r''}} 
z_1^{\tfrac{1}{2}+\tfrac{r}{2r''}}
z_2^{\tfrac{1}{2}-\tfrac{r}{2r''}}z^{-1} \\
\times&(x^{-2}z_2/z_1; x^{2r''})_\infty 
(x^{2r'}z_1/z_2; x^{2r''})_\infty 
(x^{2}z_2/z_1; x^{4})_\infty (x^{2}z_1/z_2; x^{4})_\infty \\
\times&\displaystyle\prod_{j=1}^2 
\dfrac{1}{(x^{-2}z/z_j; x^{2})_\infty 
(x^4z_j/z; x^{2})_\infty}. 
\end{array}
$$
Note that non-zero components of $F^{(2)}_{2}(S^z_1; u_1 , u_{2})$ 
on (\ref{eq:ff-final}) have poles at $z_2=x^4z_1$, which is 
consistent with the relation (\ref{eq:normalization}--\ref{eq:dual-VO21}). 

\subsection{Trigonometric limit}

Let us examine the trigonometric limit $r\rightarrow \infty$. 
The trigonometric limit of the twenty-one-vertex model, 
spin $1$ analogue of the eight-vertex model, is called 
the nineteen-vertex model. The operator algebra of 
the nineteen-vertex model can be constructed in terms of 
level $2$ irreducible highest weight representations 
of $U_q (\widehat{\mathfrak{sl}}_2)$. In what follows 
we use the same letters for both elliptic model and 
its trigonometric limit model, e.g., $S(u)$ denotes 
the $S$-matrix for both the twenty-one-vertex and the 
nineteen-vertex model. 

Unfortunately, we have no results for form factors 
of the nineteen-vertex model. Thus, let us examine 
Smirnov's axioms \cite{Smbk}, that form factors of 
integrable models should satisfy. 
Using the $S$-matrix symmetry relation (\ref{eq:R'PsiPsi}), 
the following relations should hold: 
\begin{equation}
F^{(2)}_{2}(S^z_1; u_2 , u_{1})_{
\pm 1,\mp 1} =\sum_{j=-1}^1 
F^{(2)}_{2}(S^z_1; u_1 , u_{2})_{j,-j}S(u_{1}-u_{2})_{
\mp 1,\pm 1}^{j,-j}
\label{eq:S-symmetry}
\end{equation} 
{}From (\ref{eq:S-symmetry}), we have 
\begin{equation}
{\cal F}^{(2)}_{2}(S^z_1; u_2 , u_{1})={\cal S}(u_{1}-u_{2})
{\cal F}^{(2)}_{2}(S^z_1; u_1 , u_{2}), 
\label{eq:S-symmetry-rel}
\end{equation} 
where 
$$
{\cal F}^{(2)}_{2}(S^z_1; u_1 , u_{2}):=
F^{(2)}_{2}(S^z_1; u_1 , u_{2})_{1,-1}-
F^{(2)}_{2}(S^z_1; u_1 , u_{2})_{-1,1}, 
$$
and 
$$
\begin{array}{rcl}
{\cal S}(u)&:=&S(u)_{
-1,1}^{1,-1}-S(u)_{
-1,1}^{-1,1} \\
&\rightarrow& -
\dfrac{(x\zeta^{-1}-x^{-1}\zeta)
(x^2\zeta -x^{-2}\zeta^{-1})}
{(x\zeta-x^{-1}\zeta^{-1})
(x^2\zeta^{-1} -x^{-2}\zeta)} ~~ 
(r\rightarrow \infty ). 
\end{array}
$$
Here, $\zeta =x^{u}$. 

{}From (\ref{eq:ff-final}), $F^{(2)}_{2}(S^z_1; u_1 , u_{2})_{
\pm 1,\pm 1}\rightarrow 0$ in the limit $r\rightarrow \infty$, 
which is consistent with the charge conservation under 
$U_q (\widehat{\mathfrak{sl}_2})$-symmetry. On the other hand, 
after appropriate redefinition we have 
\begin{equation}
\begin{array}{rcl}
F^{(2)}_{2}(S^z_1; u_1 , u_{2})_{
\pm 1,\mp 1} 
&\sim& \dfrac{\pm A(z_1/z)
(x^{-2}z_2/z_1; x^{4})_\infty 
(x^{2}z_1/z_2; x^{4})_\infty
\{ u_1+u_2-2u+3 \}_2}{(\zeta_1/\zeta_2 -\zeta_2/\zeta_1)
(x^2\zeta_1/\zeta_2 -x^{-2}\zeta_2/\zeta_1)} \\
&\times&\displaystyle\prod_{j=1}^2 
\dfrac{1}{(x^{-2}z/z_j; x^{2})_\infty 
(x^4z_j/z; x^{2})_\infty}, ~~~~ 
(r\rightarrow \infty )
\end{array}
\label{eq:ff-final-lim}
\end{equation}
where $A$ is some constant and $\zeta_j =x^{u_j}$\,($j=1,2$). 
Thus, our formula (\ref{eq:ff-final-lim}) satisfies 
the $S$-matrix symmetry relation (\ref{eq:S-symmetry-rel}). 

Furthermore, from the homogeneity relation (\ref{eq:homo}), 
the following cyclicity relation should hold: 
\begin{equation}
F^{(2)}_{2}(S^z_1; u_1-2 , u_{2})_{
j_1,j_2} =F^{(0)}_{2}(S^z_1; u_2 , u_{1})_{j_2,j_1}. 
\label{eq:cyclicity}
\end{equation} 
{}From (\ref{eq:cyclicity}) we have 
\begin{equation}
{\cal F}^{(2)}_{2}(S^z_1; u_1-2 , u_{2})
={\cal F}^{(2)}_{2}(S^z_1; u_2 , u_{1}). 
\label{eq:cyclicity-rel}
\end{equation} 
Here we use the relation $F^{(0)}_{2}(S^z_1; u_1 , u_{2})_{j_1,j_2}
=-F^{(2)}_{2}(S^z_1; u_1 , u_{2})_{j_1,j_2}$. 
Note that our formula (\ref{eq:ff-final-lim}) satisfies 
the cyclicity relation (\ref{eq:cyclicity-rel}) when 
$u=(u_1+u_2+3)/2$\footnote{
A similar specialization of the value of $u$ was also needed for 
Lashkevich's formula \cite{La}, the case of spin $\tfrac{1}{2}$ model. }. 

Hence we conclude that the trigonometric limit of 
the two-point form factors of the local operator 
$S^z_1$ (\ref{eq:ff-final-lim}) satisfy Smirnov's axioms 
and therefore do an appropriate $q$-difference equations of 
level $0$. 

\section{Concluding remarks} 

In this paper we have derived integral formulae for 
form factors of the twenty-one-vertex model. 
For that purpose 
we constructed the free field representations of 
the type I vertex operators $\Phi (u)_k^{k'}$ and 
the type II vertex operators $\Psi^* (u)_l^{l'}$ 
in $2\times 2$ fusion SOS model, 
the tail operators $\Lambda (u_0)_{lk}^{l'k'}$ and 
the corner transfer Hamiltonian $H^{(i)}_{l,k}$. 
Our integral formulae for form factors of 
$S^z_1$ are given by 
(\ref{eq:ff-rep}--\ref{eq:ff-rep-l}), 
which is given in terms of the $m$-fold multiple 
integrals. 

Our approach is based on some assumptions. We assumed that the vertex 
operator algebra (\ref{eq:T^Phi}--\ref{eq:T_phi}), 
(\ref{eq:T_psi}--\ref{eq:T^Psi}) and 
(\ref{eq:rho-rel}) correctly describes 
the intertwining relations between the twenty-one vertex model 
and $2\times 2$ fusion SOS model. We also assumed that 
the free field representations (\ref{eq:Bose-Lambda}), 
(\ref{eq:BoseFermi-Lambda}) 
and (\ref{eq:totalCTM}--\ref{eq:Bose/FermiCTM}) provide 
relevant representations of the vertex operator algebra. 
Using the present formalism, we can obtain 
the integral formulae for any form factors of any local 
operators in the twenty-one-vertex model, in principle. 
However, as Lashkevich said in \cite{La}, 
it is very difficult 
to obtain general formulae for form factors. In order to 
show the relevance of the present scheme, we calculated the 
simplest form factor of the local operator $S^z_1$ in subsection 4.2. 
We also show in subsection 4.3 that our form factor formulae satisfy 
an appropriate $q$-difference equations of level $0$ in 
the trigonometric limit. 

Here we wish to refer to correlation functions in 
the twenty-one-vertex model. A correlation function is 
a special example of form factors, however, 
it is not the simplest one. 
Let us recall (\ref{eq:ff-rep-l}). 
In order to calculate the form factor/correlation function 
of the spin operator $S^z_1$, we have to perform the sum 
with respect to $k_1$ and $k_2$, the state variables of the 
SOS model. There are only 
three non-zero terms $(k_1, k_2)=(k-2, k-4)$, $(k-2, k-2)$, 
$(k, k-2)$ for calculation of two-pint form factors, whereas 
there are $9(=3\times 3)$ non-zero terms\footnote{
Even after using the symmetry of $k\mapsto -k$, we should take the 
sum with respect to five terms. 
Thus, the correlation function is not the simplest example.} 
for that of correlation functions. That is why we have calculated 
not the correlation functions but the two-pint form factors of 
the spin operator $S^z_1$ as examples. 

We expect to find appropriate 
Smirnov's axiomatic structures \cite{Smbk}, 
$S$-matrix symmetry, cyclicity, and 
annihilation pole condition besides some analytic properties, 
on form factors (\ref{eq:ff-rep}--\ref{eq:ff-rep-l}). 
For that purpose, we should construct multi-point 
form factors. 
We wish to address this issue in a separate paper. 

\section*{Acknowledgements} 

I would like to thank H. Konno, 
A. Nakayashiki and M. Okado 
for discussion and their interests in the present work. 

\appendix 

\section{Appendix A ~~ Definitions of the models concerned}

\subsection{$R$-matrix of the spin $1$ analogue of the eight-vertex model}

Let $R^{(s,s')}(u)$ ($s,s'=\tfrac{1}{2}, 1, \tfrac{3}{2}, \cdots$) 
be the $R$-matrix of vertically $2s$-fold and horizontally $2s'$-fold 
fusion of $R^{(\tfrac{1}{2},\tfrac{1}{2})}(u)$, the $R$-matrix of 
the eight-vertex model. Then non-zero elements of 
$R^{(1,\tfrac{1}{2})}(u)$ are given as follows: 
\begin{equation}
\begin{array}{rcl}
R^{(\tfrac{1}{2},1)}(u)_{\pm\, \pm 1}^{\pm\, \pm 1}
&=&\dfrac{1}{\bar{\kappa}_{1,2}(u)}\dfrac{\theta_2^2 
\left( \tfrac{1}{2r} \right)\theta_2 \left( \tfrac{u}{2r} \right)}
{\theta_2^2 
\left( 0 \right)\theta_2 \left( \tfrac{2-u}{2r} \right)}, \\[6mm]
R^{(\tfrac{1}{2},1)}(u)_{\pm\, \mp 1}^{\pm\, \pm 1}
&=&\dfrac{1}{\bar{\kappa}_{1,2}(u)}\dfrac{\theta_1^2 
\left( \tfrac{1}{2r} \right)\theta_1 \left( \tfrac{u}{2r} \right)}
{\theta_2^2 
\left( 0 \right)\theta_1 \left( \tfrac{2-u}{2r} \right)}, \\[6mm]
R^{(\tfrac{1}{2},1)}(u)_{\pm\, 0}^{\pm\, 0}
&=&\dfrac{1}{\bar{\kappa}_{1,2}(u)}\dfrac{\theta_2 
\left( \tfrac{1}{r} \right)\theta_1\theta_2 \left( \tfrac{1-u}{2r} \right)}
{\theta_2 
\left( 0 \right)\theta_1\theta_2 \left( \tfrac{2-u}{2r} \right)}, 
\\[6mm]
R^{(\tfrac{1}{2},1)}(u)_{\pm\, \mp 1}^{\pm\, \mp 1}
&=&-\dfrac{1}{\bar{\kappa}_{1,2}(u)}\dfrac{\theta_2^2 
\left( \tfrac{1}{2r} \right)\theta_1 \left( \tfrac{u}{2r} \right)}
{\theta_2^2 
\left( 0 \right)\theta_1 \left( \tfrac{2-u}{2r} \right)}, \\[6mm]
R^{(\tfrac{1}{2},1)}(u)_{\pm\, \pm 1}^{\pm\, \mp 1}
&=&-\dfrac{1}{\bar{\kappa}_{1,2}(u)}\dfrac{\theta_1^1 
\left( \tfrac{1}{2r} \right)\theta_1 \left( \tfrac{u}{2r} \right)}
{\theta_2^2 
\left( 0 \right)\theta_1 \left( \tfrac{2-u}{2r} \right)}, \\[6mm]
R^{(\tfrac{1}{2},1)}(u)_{\pm\, 0}^{\mp\, \pm 1}
&=&\dfrac{1}{\bar{\kappa}_{1,2}(u)}\dfrac{\theta_1 
\left( \tfrac{1}{r} \right)\theta^2_2 \left( \tfrac{1-u}{2r} \right)}
{\theta_2 
\left( 0 \right)\theta_1\theta_2 \left( \tfrac{2-u}{2r} \right)}, \\[6mm]
R^{(\tfrac{1}{2},1)}(u)_{\pm\, 0}^{\mp\, \mp 1}
&=&-\dfrac{1}{\bar{\kappa}_{1,2}(u)}\dfrac{\theta_1 
\left( \tfrac{1}{r} \right)\theta^2_1 \left( \tfrac{1-u}{2r} \right)}
{\theta_2 
\left( 0 \right)\theta_1\theta_2 \left( \tfrac{2-u}{2r} \right)}, \\[6mm]
R^{(\tfrac{1}{2},1)}(u)_{\pm\, \mp 1}^{\mp\, 0}
&=&\dfrac{1}{\bar{\kappa}_{1,2}(u)}\dfrac{\theta_1 \theta_2 
\left( \tfrac{1}{2r} \right) \theta_2 \left( \tfrac{u}{2r} \right)}
{\theta_2^2 
\left( 0 \right)\theta_1 \left( \tfrac{2-u}{2r} \right)}, \\[6mm]
R^{(\tfrac{1}{2},1)}(u)_{\pm\, \pm 1}^{\mp\, 0}
&=&\dfrac{1}{\bar{\kappa}_{1,2}(u)}\dfrac{\theta_1 \theta_2 
\left( \tfrac{1}{2r} \right) \theta_1 \left( \tfrac{u}{2r} \right)}
{\theta_2^2 
\left( 0 \right)\theta_2 \left( \tfrac{2-u}{2r} \right)}. 
\end{array}
\label{eq:18v}
\end{equation}
where $\theta_i \left(\tfrac{u}{2r}\right)=
\theta_i \left(\tfrac{u}{2r}; \tfrac{\pi\sqrt{-1}}{2\epsilon r}\right)$, 
and 
$$
\bar{\kappa}_{1,2}(u)=(x^{-1}z)^{-\tfrac{r'}{r}} \dfrac{
(z; x^{2r})_\infty (x^{2r}z^{-1}; x^{2r})_\infty}{ 
(x^2z^{-1}; x^{2r})_\infty (x^{2r-2}z; x^{2r})_\infty}. 
$$

The case $(s,s')=(1,1)$ is of interest in the present study. 
There are twenty one non-zero elements of $R^{(1,1)}(u)$ so that 
the spin $1$ analogue of the eight-vertex model is also called 
twenty-one-vertex model. 
The explicit expressions of non-zero elements 
of $R^{(1,1)}(u)$ are given as follows: 
\begin{equation}
\begin{array}{rcl}
R^{(1,1)}(u)_{\pm1\pm1}^{\pm1\pm1}&=&\dfrac{1}{\bar{\kappa}_{2,2}(u)} 
\left( \dfrac{\theta^4_2 \left(\tfrac{1}{2r} \right) 
\theta_2 \left(\tfrac{u}{2r} \right) 
\theta_2 \left(\tfrac{1+u}{2r} \right)}{ 
\theta^4_2 \left(0 \right) 
\theta_2 \left(\tfrac{2-u}{2r} \right) 
\theta_2 \left(\tfrac{1-u}{2r} \right)} - 
\dfrac{\theta^4_1 \left(\tfrac{1}{2r} \right) 
\theta_2 \left(\tfrac{u}{2r} \right) 
\theta_1 \left(\tfrac{1+u}{2r} \right)}{ 
\theta^4_2 \left(0 \right) 
\theta_2 \left(\tfrac{2-u}{2r} \right) 
\theta_1 \left(\tfrac{1-u}{2r} \right)} \right), \\[6mm]
R^{(1,1)}(u)_{0\pm1}^{\pm1 0}&=&\dfrac{1}{\bar{\kappa}_{2,2}(u)} 
\dfrac{\theta_1\theta_2 \left(\tfrac{1}{r} \right) 
\theta^2_2 \left(\tfrac{u}{2r} \right)}{ 
\theta^2_2 \left(0 \right) 
\theta_1\theta_2 \left(\tfrac{2-u}{2r} \right)}=
R^{(1,1)}(u)^{0\pm1}_{\pm1 0}, \\[6mm]
R^{(1,1)}(u)_{\pm1 0}^{\pm1 0}&=&\dfrac{1}{\bar{\kappa}_{2,2}(u)} 
\dfrac{\theta^2_2 \left(\tfrac{1}{r} \right) 
\theta_1\theta_2 \left(\tfrac{u}{2r} \right)}{ 
\theta^2_2 \left(0 \right) 
\theta_1\theta_2 \left(\tfrac{2-u}{2r} \right)}=
R^{(1,1)}(u)^{0\pm1}_{0\pm1}, \\[6mm]
R^{(1,1)}(u)_{\pm 1\mp 1}^{\pm 1\mp 1}&=&\dfrac{1}{\bar{\kappa}_{2,2}(u)} 
\left( \dfrac{\theta^4_2 \left(\tfrac{1}{2r} \right) 
\theta_1 \left(\tfrac{u}{2r} \right) 
\theta_1 \left(\tfrac{1+u}{2r} \right)}{ 
\theta^4_2 \left(0 \right) 
\theta_1 \left(\tfrac{2-u}{2r} \right) 
\theta_1 \left(\tfrac{1-u}{2r} \right)} - 
\dfrac{\theta^4_1 \left(\tfrac{1}{2r} \right) 
\theta_1 \left(\tfrac{u}{2r} \right) 
\theta_2 \left(\tfrac{1+u}{2r} \right)}{ 
\theta^4_2 \left(0 \right) 
\theta_1 \left(\tfrac{2-u}{2r} \right) 
\theta_2 \left(\tfrac{1-u}{2r} \right)} \right), \\[6mm]
R^{(1,1)}(u)_{\pm 1\mp 1}^{\mp 1\pm 1}&=&\dfrac{1}{\bar{\kappa}_{2,2}(u)} 
\dfrac{\theta_1\theta_2 \left(\tfrac{1}{2r} \right) 
\theta_1 \left(\tfrac{1}{r} \right) 
\theta_2^3 \left(\tfrac{u}{2r} \right)}{ 
\theta^3_2 \left(0 \right) 
\theta_1 \left(\tfrac{2-u}{2r} \right) 
\theta_1\theta_2 \left(\tfrac{1-u}{2r}\right)}, \\[6mm]
R^{(1,1)}(u)_{\pm 1\mp 1}^{00}&=&-\dfrac{1}{\bar{\kappa}_{2,2}(u)} 
\dfrac{\theta_1\theta_2 \left(\tfrac{1}{r} \right) 
\theta_1\theta_2 \left(\tfrac{u}{2r} \right) 
\theta_2 \left(\tfrac{1-u}{2r} \right)}{ 
\theta^2_2 \left(0 \right) 
\theta_1\theta_2 \left(\tfrac{2-u}{2r} \right) 
\theta_1 \left(\tfrac{1-u}{2r}\right)}=
R^{(1,1)}(u)^{\pm 1\mp 1}_{00}, \\[6mm]
R^{(1,1)}(u)^{00}_{00}&=& 
\dfrac{1}{\bar{\kappa}_{2,2}(u)} 
\left( -\dfrac{\theta^2_2 \left(\tfrac{1}{r} \right) 
\theta_1\theta_2 \left(\tfrac{u}{2r} \right)}{ 
\theta^2_2 \left(0 \right) 
\theta_1 \left(\tfrac{2-u}{2r} \right)}+
\dfrac{\theta_1\theta_2 \left(\tfrac{1}{2r} \right)
\theta_1 \left(\tfrac{1}{r} \right) 
\theta^2_2 \left(\tfrac{1-u}{2r} \right)
\theta_2 \left(\tfrac{1+u}{2r} \right)}{ 
\theta^3_2 \left(0 \right) 
\theta_1\theta_2 \left(\tfrac{2-u}{2r} \right)
\theta_1 \left(\tfrac{1-u}{2r} \right)} \right. \\[6mm]
&-&\left. 
\dfrac{\theta_1\theta_2 \left(\tfrac{1}{2r} \right)
\theta_1 \left(\tfrac{1}{r} \right) 
\theta^2_1 \left(\tfrac{1-u}{2r} \right)
\theta_1 \left(\tfrac{1+u}{2r} \right)}{ 
\theta^3_2 \left(0 \right) 
\theta_1\theta_2 \left(\tfrac{2-u}{2r} \right)
\theta_2 \left(\tfrac{1-u}{2r} \right)} \right), \\[6mm]
R^{(1,1)}(u)_{\pm1\pm1}^{\mp1\mp1}&=&-\dfrac{1}{\bar{\kappa}_{2,2}(u)} 
\dfrac{\theta_1\theta_2 \left(\tfrac{1}{2r} \right)
\theta_1 \left(\tfrac{1}{r} \right) 
\theta^3_1 \left(\tfrac{u}{2r} \right)}{ 
\theta^3_2 \left(0 \right) 
\theta_2 \left(\tfrac{2-u}{2r} \right)
\theta_1\theta_2 \left(\tfrac{1-u}{2r} \right)}. 
\end{array}
\label{eq:21v}
\end{equation}
Here, 
$$
\bar{\kappa}_{2,2}(u)=z^{-\tfrac{r''}{r}} \dfrac{
(x^2z; x^{2r})_\infty (x^{2r-2}z^{-1}; x^{2r})_\infty}{ 
(x^2z^{-1}; x^{2r})_\infty (x^{2r-2}z; x^{2r})_\infty}. 
$$
Note that some of components are modified by symmetrization 
of the $R$-matrix. 

In this article we assume 
that the parameters $v$, $\epsilon$ and $r$ lie 
in the so-called principal regime (\ref{eq:principal}). 

\subsection{Boltzmann weights of $2\times 2$ fusion SOS model}

In what follows we use the following symbols: 
$$
\begin{bmatrix} u \\ m \end{bmatrix}=
\dfrac{[u]_m}{[m]_m}, ~~~~ [u]_m =[u][u-1]\cdots [u-m+1]. 
$$
Let $W_{22}$ be the Boltzmann weights of $2\times 2$ fusion SOS model, 
and let 
$$
\overline{W}_{22}
\left[ \left. \begin{array}{cc} 
c & d \\ 
b & a \end{array} \right| 
u  \right] =\bar{\kappa}^{(2,2)}(u) \begin{bmatrix} 2-u \\ 2 \end{bmatrix}
W_{22}
\left[ \left. \begin{array}{cc} 
c & d \\ 
b & a \end{array} \right| 
u  \right]
$$
be unnormalized weights. 
Then the non-zero $\overline{W}_{22}$ 
are given as follows: 
\begin{equation}
\begin{array}{l}
\overline{W}_{22}
\left[ \left. \begin{array}{cc} 
k\pm 4 & k\pm 2 \\ 
k\pm 2 & k \end{array} \right| 
u  \right] 
 =  \begin{bmatrix} 2-u \\ 2 \end{bmatrix}, 
\\[6mm]
\overline{W}_{22}
\left[ \left. \begin{array}{cc} 
k\pm 2 & k\pm 2 \\ 
k\pm 2 & k \end{array} \right| 
u  \right] 
 =  \dfrac{[1-u][k\pm 1\pm u]}
{[1][k\pm 1]}, \\[6mm] 
\overline{W}_{22}
\left[ \left. \begin{array}{cc} 
k\pm 2 & k \\ 
k & k \end{array} \right| 
u  \right] 
 =  \dfrac{[1-u][k\pm 1\mp u]}
{[1][k\pm 1]}, \\[6mm] 
\overline{W}_{22}
\left[ \left. \begin{array}{cc} 
k\pm 2 & k\pm 2 \\ 
k & k \end{array} \right| 
u  \right] 
 =  \dfrac{[k\pm 3]}
{[k\pm 1]}\begin{bmatrix} 1-u \\ 2 \end{bmatrix}, \\[6mm] 
\overline{W}_{22}
\left[ \left. \begin{array}{cc} 
k\pm 2 & k \\ 
k\pm 2 & k \end{array} \right| 
u  \right] 
 = \dfrac{[k\mp 1]}
{[k\pm 1]}\begin{bmatrix} 1-u \\ 2 \end{bmatrix}, \\[6mm] 
\overline{W}_{22}
\left[ \left. \begin{array}{cc} 
k & k\pm 2 \\ 
k\pm 2 & k \end{array} \right| 
u  \right] 
 =  
\tfrac{\begin{bmatrix} \pm k+u+1 \\ 2 \end{bmatrix}}{
\begin{bmatrix} \pm k+1 \\ 2 \end{bmatrix}}, \\[6mm] 
\overline{W}_{22}
\left[ \left. \begin{array}{cc} 
k & k\pm 2 \\ 
k\mp 2 & k \end{array} \right| 
u  \right] 
 =  
\tfrac{\begin{bmatrix} \pm k+2 \\ 2 \end{bmatrix}}{
\begin{bmatrix} \pm k \\ 2 \end{bmatrix}}
\begin{bmatrix} u+1 \\ 2 \end{bmatrix}, \\[6mm] 
\overline{W}_{22}
\left[ \left. \begin{array}{cc} 
k & k \\ 
k\pm 2 & k \end{array} \right| 
u  \right] 
 = -
\dfrac{[k\mp 1][u][k\pm u]}{
[2][k][k\pm 1]}, \\[6mm] 
\overline{W}_{22}
\left[ \left. \begin{array}{cc} 
k & k\pm 2 \\ 
k & k \end{array} \right| 
u  \right] 
 = -\dfrac{[2][k\mp 2][u][k\pm u]}{
[1]^2[k-1][k+1]}, \\[6mm] 
\overline{W}_{22}
\left[ \left. \begin{array}{cc} 
k & k \\ 
k & k \end{array} \right| 
u  \right] 
 = 
\dfrac{[k-1+u][k-u]}{
[k][k-1]}+\dfrac{[k-1][k+2]}{
[k][k+1]}\begin{bmatrix} 1-u \\ 2 \end{bmatrix}. 
\end{array}
\label{eq:BW} 
\end{equation}
Note that some of weights are modified by symmetrization 
of the Boltzmann weights. 
In this paper 
we consider so-called Regime III in the model, i.e., 
$0<u<1$. 

\subsection{Fused intertwining vectors} 

For $k'=k, k\pm 2$, let 
\begin{equation}
\begin{array}{rcl}
t(u)^k_{k'}&=&
\displaystyle\sum_{j=-1}^1 v_j 
t^j (u)^k_{k'} \\
t(u)^k_{k\pm 2}&=&\dfrac{1}{2h_1 \left(u+\tfrac{1}{2}\right)}
\begin{bmatrix} 
h_3^{(2r)}(k\mp u\pm \tfrac{3}{2})h_3^{(2r)}(k\mp u\mp \tfrac{1}{2}) \\
2h_4(1)h_4(k\mp u\pm \tfrac{1}{2}) \\
h_4^{(2r)}(k\mp u\pm \tfrac{3}{2})h_4^{(2r)}(k\mp u\mp \tfrac{1}{2})
\end{bmatrix}, \\[10mm] 
t(u)^k_{k}&=&\dfrac{1}{2h_1 \left(u+\tfrac{1}{2}\right)}
\begin{bmatrix} 
h_3^{(2r)}(k-u-\tfrac{1}{2})h_3^{(2r)}(k+u+\tfrac{1}{2}) \\
2h_4(k)h_4(u+\tfrac{1}{2}) \\
h_4^{(2r)}(k-u-\tfrac{1}{2})h_4^{(2r)}(k+u+\tfrac{1}{2}) 
\end{bmatrix}. 
\end{array}
\label{eq:int-vec-1}
\end{equation}
Then the following relation holds: 
\begin{equation}
R^{(1,1)}(u_1-u_2)t (u_1)_a^d\otimes t (u_2)_d^c=
\sum_{b} t(u_1)_b^c \otimes t (u_2)_a^b 
W_{22}\left[ \left. 
\begin{array}{cc} c & d \\ b & a \end{array} \right| 
u_1 -u_2 \right]. 
\label{eq:Rtt=Wtt1}
\end{equation}

The dual intertwining vectors are given as follows: 
\begin{equation}
\begin{array}{rcl}
t^*(u)_k^{k'}&=&
\displaystyle\sum_{j=-1}^1 v^{*j} 
t^*_j (u)_k^{k'} \\
t^*(u)_k^{k\pm 2}&=&\dfrac{\begin{bmatrix} 
h_4^{(2r)}{}^2(k\pm u\pm\tfrac{1}{2}), & 
-h_3^{(2r)}h_4^{(2r)}(k\pm u\pm\tfrac{1}{2}), & 
h_3^{(2r)}{}^2(k\pm u\pm\tfrac{1}{2})
\end{bmatrix}}
{2h_1 \left(u-\tfrac{1}{2}\right)h_1 (k) h_1(k\pm 1)} \\
t^*_1(u)_k^{k}&=&-\dfrac{h_4^{(2r)}(k+u+\tfrac{1}{2})
h_4^{(2r)}(k-u+\tfrac{3}{2})}{
2h_1 \left(u-\tfrac{1}{2}\right)h_1(k) h_1 (k+1)}-
\dfrac{h_4^{(2r)}(k-u-\tfrac{1}{2})
h_4^{(2r)}(k+u-\tfrac{3}{2})}{
2h_1 \left(u-\tfrac{1}{2}\right)h_1(k) h_1 (k-1)}, \\
t^*_0(u)_k^{k}&=&\dfrac{h_4(u-\tfrac{1}{2})(h_4(k+1)+h_4(k-1))}{
2h_1 \left(u-\tfrac{1}{2}\right)h_1(k)}\left( 
\dfrac{h_4(k+1)}{h_1(k+1)}+\dfrac{h_4(k-1)}{h_1(k-1)} \right), \\
t^*_{-1}(u)_k^{k}&=&-\dfrac{h_3^{(2r)}(k+u+\tfrac{1}{2})
h_3^{(2r)}(k-u+\tfrac{3}{2})}{
2h_1 \left(u-\tfrac{1}{2}\right)h_1(k) h_1 (k+1)}-
\dfrac{h_3^{(2r)}(k-u-\tfrac{1}{2})
h_3^{(2r)}(k+u-\tfrac{3}{2})}{
2h_1 \left(u-\tfrac{1}{2}\right)h_1(k) h_1 (k-1)}. 
\end{array}
\label{eq:dual-int-vec-1}
\end{equation}

The intertwining vectors and their dual vectors satisfy 
the following inversion relations: 
\begin{equation}
\sum_{j =-1}^1 t_j^*  (u)^{k'}_{k}
t^j (u)^{k}_{k''} =\delta_{k''}^{k'}, ~~~~ 
\sum_{k'\sim k} 
t^j (u)^{k}_{k'} 
t_{j'}^* (u)^{k'}_{k} =
\delta^j_{j'}. \label{eq:inv-1}
\end{equation}
Then the following relation holds: 
\begin{equation}
t^*(u_{1})^{b}_{c}\otimes t^*(u_{2})^{a}_{b}
R^{(1,1)}(u_{1}-u_2 )=
\displaystyle\sum_{d} 
W_{22}\left[ \left. \begin{array}{cc} 
c & d \\ b & a \end{array} \right| u_{1}-u_2 \right]
t^*(u_{1} )^{a}_{d}\otimes t^*(u_{2} )^{d}_{c}. 
\label{eq:dSOS_2}
\end{equation}

The explicit expressions of the 
$L$-operators defined by (\ref{eq:Lop}) are given as follows: 
\begin{equation}
\begin{array}{rcl}
L\left[  \left. \begin{array}{cc} k' & k'\mp 2 \\
k & k\mp 2 \end{array} \right| u_0 \right] &=&
\dfrac{\begin{bmatrix} \pm\frac{k+k'}{2} \\ 2 \end{bmatrix}
\begin{bmatrix} u_0 \pm\frac{k-k'\pm 1}{2} \\ 2 \end{bmatrix}}
{\begin{bmatrix} \pm k \\ 2 \end{bmatrix}
\begin{bmatrix} u_0+\frac{1}{2} \\ 2 \end{bmatrix}}, \\
L\left[  \left. \begin{array}{cc} k' & k'\pm 2 \\
k & k\mp 2 \end{array} \right| u_0 \right] &=&
\dfrac{\begin{bmatrix} \pm\frac{k-k'}{2} \\ 2 \end{bmatrix}
\begin{bmatrix} u_0 \pm\frac{k+k'\pm 1}{2} \\ 2 \end{bmatrix}}
{\begin{bmatrix} \pm k \\ 2 \end{bmatrix}
\begin{bmatrix} u_0+\frac{1}{2} \\ 2 \end{bmatrix}}, \\
L\left[  \left. \begin{array}{cc} k' & k'\pm 2 \\
k & k \end{array} \right| u_0 \right] &=&
\dfrac{[\frac{k+k'}{2}][\frac{k-k'}{2}]
[u_0 \pm\frac{k+k'\pm 1}{2}] 
[u_0 \pm\frac{k'-k\pm 1}{2}]}
{[k+1][k-1][u_0+\frac{1}{2}][u_0-\frac{1}{2}]}\dfrac{[2]}{[1]}, \\
L\left[  \left. \begin{array}{cc} k' & k' \\
k & k\pm 2 \end{array} \right| u_0 \right] &=&
\dfrac{[\frac{k+k'}{2}][\frac{k-k'}{2}]
[u_0 \mp\frac{k+k'\pm 1}{2}] 
[u_0 \mp\frac{k-k'\pm 1}{2}]}
{[k][k\pm 1][u_0+\frac{1}{2}][u_0-\frac{1}{2}]}, \\
L\left[  \left. \begin{array}{cc} k' & k' \\
k & k \end{array} \right| u_0 \right] &=&
\dfrac{[\frac{k+k'}{2}][\frac{k+k'}{2}-1]
[u_0 +\frac{k-k'-1}{2}][u_0 -\frac{k-k'-1}{2}]}
{[k][k-1][u_0+\frac{1}{2}][u_0-\frac{1}{2}]} \\
&+&\dfrac{[\frac{k-k'}{2}][\frac{k-k'}{2}+1]
[u_0 +\frac{k+k'+1}{2}][u_0 -\frac{k+k'+1}{2}]}
{[k][k+1][u_0+\frac{1}{2}][u_0-\frac{1}{2}]}. 
\label{eq:L-op-exp}
\end{array}
\end{equation}

\section{Appendix B ~~ OPE formulae and commutation relations}

In this Appendix we list some useful formulae for 
the basic operators. In what follows we denote 
$z=x^{2u}$, $w=x^{2v}$. 

First, useful OPE formulae are: 
\begin{eqnarray}
\Phi_1(u)\Phi_1(v)&=& z^{\frac{r''}{r}} 
\dfrac{(x^2w/z; x^{2r})_\infty}{(x^{2r'}w/z; x^{2r})_\infty}
:\Phi_1(u)\Phi_1(v):, \label{eq:PhiPhi-OPE} \\
\Phi_1(u)A(v)&=& z^{-\frac{r''}{r}} 
\dfrac{(x^{2r'}w/z; x^{2r})_\infty}{(x^2w/z; x^{2r})_\infty}
:\Phi_1(u)A(v):, \label{eq:PhiA-OPE} \\
A(v)\Phi_1(u)&=& w^{-\frac{r''}{r}} 
\dfrac{(x^{2r'}z/w; x^{2r})_\infty}{(x^2z/w; x^{2r})_\infty}
:A(v)\Phi_1(u):, \label{eq:APhi-OPE} \\
\widehat{A}(u)\widehat{A}(v)&=& z^{\frac{r''}{r}} 
\dfrac{(x^2w/z; x^{2r})_\infty}{(x^{2r'}w/z; x^{2r})_\infty}
:\widehat{A}(u)\widehat{A}(v):, \label{eq:A^-OPE} \\ 
A(u)A(v)&=& z^{\frac{r''}{r}} 
\dfrac{(x^2w/z; x^{2r})_\infty}{(x^{2r'}w/z; x^{2r})_\infty}
\left(:A(u)A(v):+ f(z,w) :\widehat{A}(u)\widehat{A}(v):\right), 
\label{eq:AA-OPE} \\
\Psi^*_1(u)\Psi^*_1(v)&=& z^{\frac{r}{r''}} 
\dfrac{(x^{-2}w/z; x^{2r''})_\infty}{(x^{2r'}w/z; x^{2r''})_\infty}
:\Psi^*_1(u)\Psi^*_1(v):, \label{eq:PsiPsi-OPE} \\
\Psi^*_1(u)B(v)&=& z^{-\frac{r}{r''}} 
\dfrac{(x^{2r'}w/z; x^{2r''})_\infty}{(x^{-2}w/z; x^{2r''})_\infty}
:\Psi^*_1(u)B(v):, \label{eq:PsiB-OPE} \\
B(v)\Psi^*_1(u)&=& w^{-\frac{r}{r''}} 
\dfrac{(x^{2r'}z/w; x^{2r''})_\infty}{(x^{-2}z/w; x^{2r''})_\infty}
:\Psi^*_1(u)B(v):, \label{eq:BPsi-OPE} \\
\widehat{B}(u)\widehat{B}(v)&=& z^{\frac{r}{r''}} 
\dfrac{(x^{-2}w/z; x^{2r''})_\infty}{(x^{2r'}w/z; x^{2r''})_\infty}
:\widehat{B}(u)\widehat{B}(v):, \label{eq:B^-OPE} \\ 
B(u)B(v)&=& z^{\frac{r}{r''}} 
\dfrac{(x^{-2}w/z; x^{2r''})_\infty}{(x^{2r'}w/z; x^{2r''})_\infty}
\left(:B(u)B(v):+ f(z,w) :\widehat{B}(u)\widehat{B}(v):\right), \\
\label{eq:BB-OPE} 
\Phi_1(u)\Psi^*_1(v)&=& \dfrac{1}{z}\dfrac{1}{1-w/z} 
:\Phi_1(u)\Psi^*_1(v):=-\Psi^*_1(v)\Phi_1(u), \label{eq:PhiPsi-OPE} \\ 
\Phi_1(u)B(v)&=& z\left(1-\dfrac{w}{z} \right)
:\Phi_1(u)B(v):=-B(v)\Phi_1(u), \label{eq:PhiB-OPE} \\
\Psi^*_1(u)A(v)&=& z\left(1-\dfrac{w}{z} \right)
:\Psi^*_1(u)A(v):=-A(v)\Psi_1(u), \label{eq:PsiA-OPE} \\
A(u)B(v)&=& \dfrac{1}{z}\dfrac{1}{1-w/z} 
\left(:A(u)B(v):+ f(z,w) :\widehat{A}(u)\widehat{B}(v):\right), 
\label{eq:AB-OPE} \\
B(v)A(u)&=& \dfrac{1}{w}\dfrac{1}{1-z/w} 
\left(:B(v)A(u):+ f(w,z) :\widehat{B}(v)\widehat{A}(u):\right), 
\label{eq:BA-OPE} \\
W(v)\Phi_1(u)&=&w^{\tfrac{2}{r}}\dfrac{(x^{r-2}z/w; x^{2r})_\infty}{
(x^{r+2}z/w; x^{2r})_\infty}:W(v)\Phi_1(u):, \label{eq:WPhi-OPE} \\
W(v)A(v')&=&w^{-\tfrac{2}{r}}\dfrac{(x^{r+2}w'/w; x^{2r})_\infty}{
(x^{r-2}w'/w; x^{2r})_\infty}:W(v)A(v'):, \label{eq:WA-OPE} \\
W(v)\Psi^*_1(u)&=&w^{-\tfrac{2}{r''}}\dfrac{(x^{r}z/w; x^{2r''})_\infty}{
(x^{r-4}z/w; x^{2r''})_\infty}:W(v)\Phi_1(u):, \label{eq:WPsi-OPE} \\
W(v)B(v')&=&w^{\tfrac{2}{r''}}\dfrac{(x^{r-4}w'/w; x^{2r''})_\infty}{
(x^{r}w'/w; x^{2r''})_\infty}:W(v)B(v'):. \label{eq:WB-OPE} 
\end{eqnarray}
Here $\widehat{A}(v)$ and $\widehat{B}(v)$ denote the fermion contraction
$$
\widehat{A}(v)=w^{\frac{r''}{2r}}:\exp \left( 
\displaystyle\sum_{m\neq 0} 
\frac{\beta_m}{m}w^{-m}\right):e^{-\alpha} 
w^{\frac{1}{2}L-\frac{r''}{2r}K}, 
$$
$$
\widehat{B}(v)=w^{\frac{r}{2r''}}:\exp \left( -
\displaystyle\sum_{m\neq 0} 
\dfrac{[rm]_x}{[r''m]_x}\frac{\beta_m}{m}w^{-m}\right):e^{-\beta} 
w^{-\frac{r}{2r''}L+\frac{1}{2}K}, 
$$
and 
\begin{equation}
f(z,w)=\dfrac{1}{x+x^{-1}}\displaystyle\sum_{m>0} 
\left( \left(\dfrac{x^2w}{z}\right)^m+
\left(\dfrac{x^{-2}w}{z}\right)^m \right). 
\label{eq:f(z,w)}
\end{equation}
From these, we obtain the following 
commutation relations: 
\begin{eqnarray}
\Phi_1(u)\Phi_1(v)&=&\dfrac{[v-u+1]}{[u-v+1]}
\Phi_1(v)\Phi_1(u), \\
A(v)\Phi_1(u)&=&\dfrac{[v-u+1]}{[u-v+1]}
\Phi_1(u)A(v), \\
\left[ u-v+1 \right] A(u)A(v)&=&\left[ u-v-1\right] A(v)A(u), 
\label{eq:AA-com} \\ 
\Psi^*_1(u)\Psi^*_1(v)&=&\dfrac{[v-u-1]''}{[u-v-1]''}
\Psi^*_1(v)\Psi^*_1(u), \\
B(v)\Psi^*_1(u)&=&\dfrac{[v-u-1]''}{[u-v-1]''}
\Psi^*_1(u)B(v), \\
\left[ u-v-1 \right]'' B(u)B(v)&=&\left[ u-v+1\right]'' B(v)B(u), \\
\Phi_1(u)\Psi^*_1(v)&=& -\Psi^*_1(v)\Phi_1(u), \\
\Phi_1(u)B(v)&=& B(v)\Phi_1(u), \\
\Psi^*_1(u)A(v)&=& A(v)\Psi^*_1(u), \\
{[}A(u), B(v){]}&=& \dfrac{1}{(x+x^{-1})(z-w)}:\widehat{A}(u)\widehat{B}(v): 
\left( \delta\left(\dfrac{x^2w}{z}\right)+
\delta\left(\dfrac{x^{-2}w}{z}\right) \right), 
\label{eq:AB-com} \\
W(v)\Phi_1(u)&=&\dfrac{[u-v+\tfrac{r''}{2}]}{[v-u+\tfrac{r''}{2}]}
\Phi_1(u)W(v), \\
W(v)A(v')&=&\dfrac{[v-v'+\tfrac{r''}{2}]}{[v'-v+\tfrac{r''}{2}]}
A(v')W(v), \label{eq:WA-comm} \\
W(v)\Psi^*_1(u)&=&\dfrac{[u-v+\tfrac{r}{2}]''}{[v-u+\tfrac{r}{2}]''}
\Psi\*_1(u)W(v), \\
W(v)B(v')&=&\dfrac{[v-v'+\tfrac{r}{2}]''}{[v'-v+\tfrac{r}{2}]''}
B(v')W(v). \label{eq:WB-comm}
\end{eqnarray}

\section{Appendix C ~~ Free field representations of 
$\Lambda (u_0)_{lk}^{l'k'}$}

Consider the LHS of (\ref{eq:Lambda-Psi2}) with $k'=k-2$. 
\begin{equation}
\begin{array}{rcl}
[\Psi^*(u)_l^{l-2}, \Lambda (u_0 )^{lk-2}_{lk}]&=&c_1 \displaystyle
\oint_{C'}\dfrac{dw}{2\pi\sqrt{-1}w} 
\oint_{C}\dfrac{dw'}{2\pi\sqrt{-1}w'} [B(v), A(v')] 
\Psi^*_1 (u) \\
&\times& \dfrac{[v-u+l]''}{[v-u-1]''} 
\dfrac{[v'-u_0-\tfrac{1}{2}-k]}{[v'-u_0-\tfrac{3}{2}]}, 
\end{array}
\label{eq:Lambda-Psi2''}
\end{equation}
where 
$$
c_s=(-1)^s\dfrac{[s+1][k-2s][k-s+1]}{[1][k][k+1]}. 
$$
Using (\ref{eq:AB-com}), the integral with respect to 
$w'$ on (\ref{eq:Lambda-Psi2''}) 
can be evaluated by the substitution $w'=x^{\pm 2}w$. 
The result is as follows: 
\begin{equation}
\begin{array}{cl}
&[\Psi^*(u)_l^l, \Lambda (u_0 )^{lk-2}_{lk}] \\
=&\displaystyle\dfrac{c_1}{x^{-2}-x^{2}} 
\oint_{C'}\dfrac{dw}{2\pi\sqrt{-1}w} \left( 
F\left( v-\tfrac{r''}{2} \right) W\left( v-\tfrac{r''}{2} \right)
-F\left( v+\tfrac{r''}{2} \right) W\left( v+\tfrac{r''}{2} \right)
\right) 
\Psi^*_1 (u) \\
=& \displaystyle\dfrac{c_1}{x^{-2}-x^{2}} 
\left( \oint_{x^{-r''}C'}\dfrac{dw}{2\pi\sqrt{-1}w} 
-\oint_{x^{r''}C'}\dfrac{dw}{2\pi\sqrt{-1}w} \right) F(v)W(v)\Psi^*_1 (u), 
\end{array}
\label{eq:Lambda-Psi2'''}
\end{equation}
where 
$$
F(v):= \dfrac{[v-u-\tfrac{r''}{2}+l]''}{[v-u-\tfrac{r}{2}]''} 
\dfrac{[v-u_0+\tfrac{r-1}{2}-k]}{[v-u_0+\tfrac{r-3}{2}]}. 
$$
The integral with respect to 
$w$ on (\ref{eq:Lambda-Psi2'''}) 
can be evaluated by the residue at $w=x^{r}z$ and 
$w=x^{3-r}z_0$. The former residue vanishes because of 
(\ref{eq:WPsi-OPE}). Thus, from (\ref{eq:Lambda-Psi2'}) we have 
$$
\dfrac{c_1}{x^{-2}-x^{2}} 
\dfrac{[u_0-u+\tfrac{1}{2}+l]''}{[u_0-u-\tfrac{1}{2}]''} 
\dfrac{[1-k]}{\partial [0]}W_-(u_0)=\Lambda (u_0 )^{l\,k-2}_{l+2\,k}
\dfrac{[1]''[u_0+\Delta u-u+l+\tfrac{1}{2}]''}{[l+2]''
[u_0+\Delta u-u-\tfrac{1}{2}]''}. 
$$
Hence we conclude that $\Delta u=0$ and 
\begin{equation}
\Lambda (u_0 )^{l\,k-2}_{l+2\,k}=
\dfrac{[l+2]''}{[1]''}\dfrac{[2][k-1][k-2]}{(x^{-2}-x^2)
\partial [0][1][k+1]}W_-(u_0). 
\end{equation}
Let us summarize the result as follows: 
\begin{equation}
\begin{array}{rcl}
[Y^*(u), X (u_0+\tfrac{1}{2} )]\Psi^*_1 (u)&=&
\displaystyle\dfrac{1}{x^{-2}-x^2}W_-(u_0) \Psi^*_1 (u)
\dfrac{[u_0-u+\tfrac{1}{2}+L]''}{[u_0-u-\tfrac{1}{2}]''} 
\dfrac{[1-K]}{\partial [0]}. 
\end{array}
\label{eq:Y*X-comm}
\end{equation}

Consider the LHS of (\ref{eq:Lambda-Psi2}) with $k'=k-2s$. 
\begin{equation}
\begin{array}{rcl}
[\Psi^*(u)_l^l, \Lambda (u_0 )^{lk'}_{lk}]&=&c_s \displaystyle
\oint_{C'}\dfrac{dw}{2\pi\sqrt{-1}w} \prod_{j=1}^s
\oint_{C}\dfrac{dw'_j}{2\pi\sqrt{-1}w'_j} 
[B(v), A(v'_s)\cdots A(v'_1)] 
\Psi^*_1 (u) \\
&\times& 
\dfrac{[v-u+l]''}{[v-u-1]''} 
\dfrac{[v'_s-u_0-\tfrac{1}{2}-(k-2s+2)]}{[v'_s-u_0-\tfrac{3}{2}]}\cdots 
\dfrac{[v'_1-u_0-\tfrac{1}{2}-k]}{[v'_1-u_0-\tfrac{3}{2}]} \\
&=& c_s \displaystyle\dfrac{[s]!}{s![1]^s} 
\oint_{C'}\dfrac{dw}{2\pi\sqrt{-1}w} \prod_{j=1}^s
\oint_{C}\dfrac{dw'_j}{2\pi\sqrt{-1}w'_j} 
[B(v), A(v'_s)\cdots A(v'_1)] 
\Psi^*_1 (u) \\
&\times& \displaystyle\dfrac{[v-u+l]''}{[v-u-1]''} 
\prod_{i<j}^s \dfrac{[v'_i-v'_j]}{[v'_i-v'_j+1]} 
\prod_{i=1}^s \dfrac{[v'_i-u_0-\tfrac{1}{2}-(k-s+1)]}
{[v'_i-u_0-\tfrac{3}{2}]}. 
\end{array}
\label{eq:Lambda-Psi2_s}
\end{equation}
Using (\ref{eq:Y*X-comm}) and the commutation relation 
(\ref{eq:WA-comm}) we have 
\begin{equation}
\begin{array}{cl}
&[\Psi^*(u)_l^l, \Lambda (u_0 )^{lk'}_{lk}] \\
=&\dfrac{c_s}{x^{-2}-x^2} \dfrac{[s]!}{(s-1)![1]^s} 
\displaystyle
\prod_{j=1}^{s-1}
\oint_{C}\dfrac{dw'_j}{2\pi\sqrt{-1}w'_j}
W_-(u_0) A(v'_{s-1})\cdots A(v'_1) 
\Psi^*_1 (u) \\
\times& \displaystyle
\dfrac{[u_0-u+\tfrac{1}{2}+l]''}{[u_0-u-\tfrac{1}{2}]''} 
\dfrac{[s-k]}{\partial [0]}
\prod_{i<j}^{s-1} \dfrac{[v'_i-v'_j]}{[v'_i-v'_j+1]} 
\prod_{i=1}^{s-1} \dfrac{[v'_i-u_0-\tfrac{1}{2}-(k-s+1)]}
{[v'_i-u_0-\tfrac{1}{2}]} \\
=& \dfrac{c_s}{x^{-2}-x^2} 
\dfrac{[s][s-k]}{\partial [0][1]}
\dfrac{[u_0-u+\tfrac{1}{2}+l]''}{[u_0-u-\tfrac{1}{2}]''} 
W_-(u_0) X(u_0-\tfrac{1}{2})^{s-1}. 
\end{array}
\label{eq:Lambda-Psi2_s'}
\end{equation}
{}From (\ref{eq:Lambda-Psi2'}) with $\Delta u=0$ 
and (\ref{eq:Lambda-Psi2_s'}) 
we obtain (\ref{eq:BoseFermi-Lambda1}). 

Consider the LHS of (\ref{eq:Lambda-Psi3}) with $k'=k-2s$. 
\begin{equation}
\begin{array}{cl}
&[\Psi^*(u)_l^{l-2}, \Lambda (u_0 )^{lk'}_{lk}] \\
=&c_s \displaystyle\prod_{a=1}^2
\oint_{C'}\dfrac{dw_a}{2\pi\sqrt{-1}w_a} \prod_{j=1}^s
\oint_{C}\dfrac{dw'_j}{2\pi\sqrt{-1}w'_j} 
[B(v_1)B(v_2), A(v'_s)\cdots A(v'_1)] 
\Psi^*_1 (u) \\
\times& 
\dfrac{[v_1-u+l-2]''}{[v_1-u-1]''} 
\dfrac{[v_2-u+l]''}{[v_2-u-1]''} 
\dfrac{[v'_s-u_0-\tfrac{1}{2}-(k-2s+2)]}{[v'_s-u_0-\tfrac{3}{2}]}\cdots 
\dfrac{[v'_1-u_0-\tfrac{1}{2}-k]}{[v'_1-u_0-\tfrac{3}{2}]} \\
=& c_s \displaystyle\dfrac{[s]!}{s![1]^s} \dfrac{[2]''}{2[1]''}
\prod_{a=1}^2\oint_{C'}\dfrac{dw_a}{2\pi\sqrt{-1}w_a} \prod_{j=1}^s
\oint_{C}\dfrac{dw'_j}{2\pi\sqrt{-1}w'_j} 
[B(v_1)B(v_2), A(v'_s)\cdots A(v'_1)] 
\Psi^*_1 (u) \\
\times& \displaystyle\dfrac{[v_1-v_2]''}{[v_1-v_2+1]''} 
\prod_{a=1}^2 \dfrac{[v_a-u+l-1]''}{[v_a-u-1]''} 
\prod_{i<j}^s \dfrac{[v'_i-v'_j]}{[v'_i-v'_j+1]} 
\prod_{i=1}^s \dfrac{[v'_i-u_0-\tfrac{1}{2}-(k-s+1)]}
{[v'_i-u_0-\tfrac{3}{2}]}. 
\end{array}
\label{eq:Lambda-Psi3_s}
\end{equation}
Using (\ref{eq:Y*X-comm}) and the commutation relations 
(\ref{eq:WA-comm}, \ref{eq:WB-comm}) we have 
\begin{equation}
\begin{array}{cl}
&[\Psi^*(u)_l^{l-2}, \Lambda (u_0 )^{lk'}_{lk}]=
\dfrac{c_s}{x^{-2}-x^2} \dfrac{[s]!}{(s-1)![1]^s} 
\dfrac{[2]''}{[1]''} 
\dfrac{[u_0-u-\tfrac{1}{2}+l]''}{[u_0-u-\tfrac{1}{2}]''} 
\dfrac{[s-k]}{\partial [0]} \\
\times& \displaystyle
\oint_{C'}\dfrac{dw_1}{2\pi\sqrt{-1}w_1} \prod_{j=1}^{s-1}
\oint_{C}\dfrac{dw'_j}{2\pi\sqrt{-1}w'_j} B(v_1)
W_-(u_0) A(v'_{s-1})\cdots A(v'_1) 
\Psi^*_1 (u) \\
\times& \displaystyle
\dfrac{[v_1-u_0-\tfrac{1}{2}]''}{[v_1-u_0+\tfrac{1}{2}]''} 
\dfrac{[v_1-u+l-1]''}{[v_1-u-1]''} 
\prod_{i<j}^{s-1} \dfrac{[v'_i-v'_j]}{[v'_i-v'_j+1]} 
\prod_{i=1}^{s-1} \dfrac{[v'_i-u_0-\tfrac{1}{2}-(k-s+1)]}
{[v'_i-u_0-\tfrac{1}{2}]}. 
\end{array}
\label{eq:Lambda-Psi3_s'}
\end{equation}
Using (\ref{eq:Lambda-Psi3_s'}) and 
$$
\begin{array}{rcl}
L''\left[ \left. \begin{array}{cc} l-2 & l \\
l & l \end{array} \right| u_0-u \right] 
&=&\dfrac{[2]''[u_0-u+l-\tfrac{1}{2}]''}{[l+1]''
[u_0-u+\tfrac{1}{2}]''}, \\[5mm]
L''\left[ \left. \begin{array}{cc} l-2 & l \\
l+2 & l \end{array} \right| u_0-u \right]&=& 
\dfrac{[1]''[2]''[u_0-u+l-\tfrac{1}{2}]''[u_0-u+l+\tfrac{1}{2}]''}
{[l+1]''[l+2]''[u_0-u-\tfrac{1}{2}]''[u_0-u+\tfrac{1}{2}]''}, 
\end{array}
$$
the relation (\ref{eq:Lambda-Psi3}) reduces to 
\begin{equation}
\begin{array}{cl}
& \dfrac{[1]''[u_0-u+l+\tfrac{1}{2}]''}
{[l+2]''[u_0-u+\tfrac{1}{2}]''}\Lambda(u_0)_{l+2\,k}^{l-2\,k'} \\
=& \dfrac{c_s}{x^{-2}-x^2} \dfrac{[s]}{[1]} 
\dfrac{[l+1]''}{[1]''}
\dfrac{[s-k]}{\partial [0]} \displaystyle
\oint_{C'}\dfrac{dw_1}{2\pi\sqrt{-1}w_1} 
\dfrac{[v_1-u_0-\tfrac{1}{2}]''}{[v_1-u_0+\tfrac{1}{2}]''} 
\dfrac{[v_1-u+l-1]''}{[v_1-u-1]''} \\
\times & B(v_1)
W_-(u_0) X(u_0-\tfrac{1}{2})^{s-1} -
\dfrac{[u_0-u-\tfrac{1}{2}]''}{[u_0-u+\tfrac{1}{2}]''}
\Lambda(u_0)_{lk}^{l-2\,k'}Y^*(u). 
\end{array}
\label{eq:Lambda-Psi3_s''}
\end{equation}
{}Using the commutation relation (\ref{eq:WB-comm}) and 
the addition theorem 
\begin{equation*}
\begin{array}{cl}
&\dfrac{[l+1]''}{[1]''}
\dfrac{[v_1-u_0-\tfrac{1}{2}]''}{[v_1-u_0+\tfrac{1}{2}]''} 
\dfrac{[v_1-u+l-1]''}{[v_1-u-1]''}-\dfrac{[l]''}{[1]''}
\dfrac{[u_0-u-\tfrac{1}{2}]''}{[u_0-u+\tfrac{1}{2}]''}
\dfrac{[v_1-u+l]''}{[v_1-u-1]''}
\dfrac{[v_1-u_0-\tfrac{3}{2}]''}{[v_1-u_0+\tfrac{1}{2}]''} \\
=&
\dfrac{[u_0-u+l+\tfrac{1}{2}]''}{[u_0-u+\tfrac{1}{2}]''}
\dfrac{[v_1-u_0-\tfrac{1}{2}+l]''}{[v_1-u_0+\tfrac{1}{2}]''}, 
\end{array}
\end{equation*}
the relation (\ref{eq:Lambda-Psi3_s''}) reduces to 
(\ref{eq:BoseFermi-Lambda2}). 

Repeating the similar procedures we can derive 
the general expression (\ref{eq:BoseFermi-Lambda}).


\begin{thebibliography}{99}
\bibitem{ESM}Baxter R J: {\it Exactly Solved Models 
in Statistical Mechanics}, 
Academic Press, London, 1982.
\bibitem{JMbk} Jimbo M and Miwa T: 
{\it Algebraic analysis of solvable lattice models}, 
CBMS Regional Conferences Series in Mathematics Vol {\bf 85}; 
AMS: Providence, RI, 1994. 
\bibitem{LaP}Lashkevich M and Pugai Ya: 
Free field construction for correlation 
functions of the eight-vertex model, 
{\it Nucl. Phys.} {\bf B516} 623--651, 1998. 
\bibitem{LuP} Lukyanov S and Pugai Ya: 
Multi-point local height probabilities in the 
integrable RSOS model, 
{\it Nucl. Phys.} {\bf B473}[FS] 631--658, 1996. 
\bibitem{La}Lashkevich M: Free field construction 
for the eight-vertex model: representation for form factors. 
{\it Nucl. Phys.} {\bf B621} 587--621, 2002. 
\bibitem{KKW}Kojima T, Konno H and Weston R: 
The vertex-face correspondence and correlation functions of 
the fusion eight-vertex model I: The general formalism. 
{\it Nucl. Phys.} {\bf B720} [FS] 348--398, 2005. 
\bibitem{Bela}Belavin A A: Dynamical symmetry 
of integrable quantum systems, {\it Nucl. Phys.} 
{\bf B180}[FS2] 189--200, 1981. 
\bibitem{Bel-corr}Quano Y.-H: 
A vertex operator approach for correlation functions of Belavin's 
$(\mathbb{Z}/n\mathbb{Z})$-symmetric model, 
{\it J. Phys. A: Math. Theor.} {\bf 42} 165211, 20 pages, 2009. 
\bibitem{Bel-form}Quano Y.-H: 
Vertex operator approach for form factors of Belavin's 
$(\mathbb{Z}/n\mathbb{Z})$-symmetric model, 
{\it J. Phys. A: Math. Theor.} {\bf 43} 
085202 doi:10.1088/1751-8113/43/8/085202, 23 pages, 2010. 
\bibitem{DJMO}Date E, Jimbo M, Miki K and Okado M: 
Mean staggered polarization for the higher spin analog 
of the $6$-vertex model, 
{\it Int. J. Mod. Phys.} {\bf A7} Suppl.1A 151--163, 1992. 
\bibitem{Idz}Idzumi M: Correlation functions of 
the spin-1 analog of the XXZ model, hep-th/9307129, 1993; 
Level 2 irreducible representations of $U_q (\widehat{sl_2})$, 
vertex operators, and their correlations, 
{\it Int. J. Mod. Phys.} {\bf A9} 4449--4484, 1994.
\bibitem{Reshe}Reshetikhin N Yu: 
$S$-matrices in integrable models of isotropic magnetic chains. I, 
{\it J. Phys. Math. Gen.} {\bf A24} 3299--3309, 1991. 
\bibitem{BW}Bougourzi A H and Weston R A: 
$N$-point correlation functions of the spin-1 XXZ model, 
{\it Nucl. Phys.} {\bf B417} 439--462, 1994. 
\bibitem{8VSOS1}Baxter R J: Eight-vertex model in lattice statistics and 
one-dimensional anisotropic Heisenberg chain I. 
Some fundamental eigenvectors, 
{\it Ann. Phys.} {\bf 76} 1--24, 1973. 
\bibitem{8VSOS2}Baxter R J: Eight-vertex model in lattice statistics and 
one-dimensional anisotropic Heisenberg chain II. 
Equivalence to a generalized ice-type model, 
{\it Ann. Phys.} {\bf 76} 25--47, 1973. 
\bibitem{8VSOS3}Baxter R J: Eight-vertex model in lattice statistics and 
one-dimensional anisotropic Heisenberg chain III. 
Eigenvectors of the transfer matrix and Hamiltonian, 
{\it Ann. Phys.} {\bf 76} 48--71, 1973. 
\bibitem{BJMST}Boos H, Jimbo M, Miwa T, Smirnov F and Takeyama Y,
Traces on the Sklyanin algebra and correlation functions of 
the eight-vertex model, {\it J. Phys. A: Math. Gen} {\bf 38} 
7629--7659, 2005.
\bibitem{ABF}Andrews G E, Baxter R J and 
Forrester P J: Eight-vertex SOS model and 
generalized Rogers--Ramanujan--type identities, 
{\it J. Stat. Phys.} {\bf 35} 193--266, 1984. 
\bibitem{FusionSOS1}Date E, Jimbo M, Kuniba A, Miwa T 
and Okado M: Exactly solvable SOS models: 
Local height probabilities and theta function identities, 
{\it Nucl. Phys.} {\bf B290} 231--273, 1987. 
\bibitem{FusionSOS2}Date E, Jimbo M, Kuniba A, Miwa T 
and Okado M: Exactly solvable SOS models II: 
Proof of the star-triangle relation and combinatorial identities, 
{\it Adv. Stud. Pure Math.} {\bf 16} 17--122, 1988. 
\bibitem{DJKMO1}Date E, Jimbo M, Kuniba A, Miwa T and Okado M: 
One dimensional configuration sums in vertex models and 
affine Lie algebra characters, 
{\it Lett. Math. Phys.} {\bf 17} 69--77, 1989. 
\bibitem{DJKMO2}Date E, Jimbo M, Kuniba A, Miwa T and Okado M: 
Paths, Maya diagrams and representation of $\widehat{sl}(r,{\bf C})$, 
{\it Adv. Stud. Pure Math.} {\bf 19} 149--191, 1989. 
\bibitem{KP}Kac V G and Peterson D H: 
Infinite dimensional Lie algebra, theta-functions 
and modular forms, {\it Adv. Math.} {\bf 53} 125--264, 1984. 
\bibitem{JKOS1}Jimbo M, Konno H, Odake S and Shiraishi J: 
Quasi-Hopf twistors for elliptic quantum groups, 
{\it Transform. Groups} {\bf 4} 303--327, 1999. 
\bibitem{Ko}Konno H, An elliptic algebra 
$U_{q,p}(\widehat{\mathfrak{sl}_2})$ and the fusion RSOS 
model, {\it Commun. Math. Phys.} {\bf 195} 373--403, 1998. 
\bibitem{JKOS2}Jimbo M, Konno H, Odake S and Shiraishi J: 
Elliptic algebra $U_{q,p}(\widehat{sl}_2)$ : 
Drinfeld currents and vertex operators, 
{\it Commun. Math. Phys.}, {\bf 199} 605--647, 1999. 
\bibitem{Czech}Konno H: Free field realisation of 
the level-$2$ elliptic algebra $U_{x,p}(\widehat{\mathfrak{sl}}_2)$, 
{\it Czech. J. of Phys.} {\bf 55} 1455--1460, 2005. 
\bibitem{Annecy}Konno H: Correlation functions of the spin $1$ XYZ model, 
A talk given at the workshop 
``Recent Advances in Quantum Integrable Systems'', 
LAPTH, Annecy-le-Vieux, France, Sept. 2005. 
\bibitem{Smbk}F. A. Smirnov, {\it 
Form factors in completely integrable models of
quantum field theory}, Advanced Series in Mathematical 
Physics Vol {\bf 14}, (World Scientific, Singapore, 1992).
\end{thebibliography}
\end{document}